%% file: Article1.tex
\documentclass[traditabstract]{aa}

\usepackage{txfonts}
\usepackage[english]{babel}

\usepackage{graphicx}

\usepackage{natbib}
\usepackage{xspace}
\usepackage{enumitem} 
\usepackage{color}

\usepackage{amsmath}
\usepackage{amssymb, amsfonts}

\setlist[itemize]{noitemsep} 

\def\mhmpc{\,h^{-1}~\rm{Mpc}}
\def\Mpc{\,h^{-1}~\rm{Mpc}}

\DeclareMathOperator\erf{erf}
\providecommand{\E}[1]{\ensuremath{\mathbb{E}\left[#1\right]}}
\providecommand{\Var}[1]{\ensuremath{\mathbb{V}\left[#1\right]}}

\newcommand{\Norm}{\ensuremath{{\mathcal N}}}

\newcommand{\mat}[1]{\ensuremath{{\mathbf #1}}\xspace}
\newcommand{\nn}{\nonumber}

\newcommand{\pdf}{p.d.f\xspace}

\newcommand{\lcal}{{\mathcal L}}

\newcommand{\fcal}{{\mathcal F}}

\newcommand{\pcal}{{\mathcal P}}
\newcommand{\kvec}{{\vec k}}
\newcommand{\dif}{{\mathrm d}}

\begin{document}
\title{High-precision Monte-Carlo modelling of galaxy distribution}



\author{
Philippe Baratta\inst{1,}\inst{2}
\and Julien Bel \inst{2}
\and Stephane Plaszczynski \inst{3}
\and Anne Ealet \inst{1,}\inst{4}
}

\institute{
Aix Marseille Universit\'e, CNRS/IN2P3, CPPM, Marseille, France 
\and Aix Marseille Univ, Universit\'e de Toulon, CNRS, CPT, Marseille, France
\and LAL, Univ. Paris-Sud, CNRS/IN2P3, Universit\'e Paris-Saclay, Orsay, France
\and Institut de Physique Nucléaire de Lyon, 69622, Villeurbanne, France
}


\offprints{\mbox{J.~Bel}, \email{jbel@cpt.univ-mrs.fr} }

\abstract{
We revisit the case of fast Monte-Carlo simulations of galaxy positions
for a non-gaussian field. More precisely we address the question of
generating a 3D field with a given one-point function (as
a log-normal one, but not only)  and some power-spectrum fixed by
cosmology.
We highlight and investigate a problem that occurs when the field is filtered
and identify, for the log-normal case, a regime where it can still be used. 
However we show that the filtering is unnecessary if one takes into account
aliasing effects and finely controls the discrete sampling step.
In this way we demonstrate a sub-percent precision of all our
spectra up to the Nyquist frequency. We extend the
method to generate a full light cone evolution comparing two methods for doing
it and validate our method with a tomographic analysis. 
We investigate analytically and numerically the structure of the covariance matrices
obtained with such simulations which may be useful for future large and
deep surveys.
}


\keywords{Monte-Carlo simulations  -  large scale structures  -  trispectrum  -  2-points statistics  -  covariance matrix  -  log-normal}

\maketitle

\input{introduction.tex}
\input{section_0.tex}
\input{section_1.tex}

\input{section_2.tex}

\input{section_3.tex}
\input{conclusion.tex}


\appendix
\input{appendix_LN.tex}
\input{appendix_Mehler.tex}
\input{appendix_trispectre.tex}

\bibliographystyle{aa} 
\bibliography{refs}


\end{document}

%% file: introduction.tex
\section*{Introduction}
\label{introduction}

Fast Monte-Carlo methods  are essential tools to design analyses over large
datasets. Widely used in the Cosmological Microwave Background (CMB)
community, thanks to the high quality \verb Healpix  software \citep{Healpix:2005} 
they are less frequently used in galactic surveys where analyses often rely
on mock catalogues following a complicated and heavy process chain.
The reason is that the problem is more complex since w.r.t to the CMB
\begin{itemize}
\item the galaxy distribution follows a 3D stochastic point process,
\item the underlying continuous field is not Gaussian.
\end{itemize}

The first point, that leads to shot noise, can be accommodated
although a Monte Carlo tool cannot provide universal "window functions",
for correcting voxels effects since data do not lie on a sphere but on some complicated 3D domain.

The matter distribution field \texttt{cannot} be Gaussian.
A very simple way to see it is to note that even in the so-called "linear" regime, i.e for
scales above $\simeq 8 \mhmpc$ one measures $\sigma_8\simeq 0.8$ which
represents
the standard-deviation of the matter \textit{density contrast}
$\delta={\rho}/{\bar \rho}-1$. Would the one point distribution $P(\delta)$ follow
a Gaussian with such a standard deviation, the energy density $\rho=\bar \rho(1+\delta)$ would
become negative in about $P(\delta \le -1)= 10\%$ of the cases!  This very obvious
argument demonstrates that even in what is called the "linear regime",
the field is not Gaussian and follows some more evolved distribution.

This is a serious problem because non Gaussian fields are difficult to
characterize \citep{Adler:1981} and shooting samples following them is a Herculean task.
Cosmologists focussed essentially on the subset of fields obtained
by applying some transformation to a Gaussian one \citep{Coles:1987}.
Remarkably, in some (rare) cases the the auto-correlation function of the
transformed field has can be expressed analytically from the Gaussian one. This happens for the
log-normal (LN) field, obtained essentially by taking the
exponential of a Gaussian field \citep{Coles:1991} which largely explains the reason for its success in cosmology.

Since Hubble conjectured it in 1934 \citep{Hubble:1934} it still describes surprisingly well 
the 1-point distribution of
galaxies in the $\sigma <1$ regime \citep{Clerkin:2017}, given that it has no theoretical foundations.
A closer look, based on numerous N-body simulations, reveals it is not perfect
in particular for higher variances, thus extensions with more freedom
such as the skewed log-normal \citep{colombi94} or the Gamma expansion \citep{GF&E2000}.
More recently,  \citet{Klypin:2018}  proposed some more refined parameterisations.
One may prefer a more physical description as the one based on a large deviation principle and spherical infall model
\citep{Uhlemann:2016} that provides a fully deterministic formula for the \pdf in the mildly non linear
regime \citep{Codis:2016}.

Boltzmann codes as \verb CLASS  \citep{Blas:2011rf}, by solving numerically the perturbation equations in
the linear regime and adding some contribution from models for small scales,
predict the \texttt{matter power spectrum} for a
given cosmology. For any field, this quantity is always defined as the Fourier 
Transform of the auto-correlation function. Only in the Gaussian case
does it contain all the available information.
Then if we want to study cosmological parameters we need to provide realisations that follow some given spectrum.

In the following, we present a method for generating a matter field
(and subsequent catalogs) following any one-point function and some target power-spectrum.
Although it is similar to standard methods for generating a LN field
\citep[e.g][]{Chiang:2013,Greiner:2015,Agrawal:2017}
it is more general and solves an important issue.
The way of generating a LN distribution with a target power-spectrum by
transforming a Gaussian field is  
\textit{ill-defined} when the field is smoothed, since it requires an
input "power spectrum" with some negative parts. We will focus on
that problem and show its origin in Sect.\ref{sec:LN}.
Then we will show in  Sect.\ref{sec:sampling} how this problem can
be cured by adjusting the discretization step and including aliasing effects. 
We will use the Mehler transform to show how any form of the probability distribution function (\pdf) can be achieved
still keeping a positive input power-spectrum. We then give an
analytical expression of the general tri-spectrum and compare it to
the output of the simulations. 
In Sect.\ref{sec:sampling} we consider the production of a discrete catalog
and how the cell window function affects the result. We discuss a
linear interpolation scheme that reduces discontinuities between cells.
We also consider two methods to account for the
redshift evolution, one with the full light-cone reconstruction and
the other evolving the perturbation, which will be compared.
To qualify our catalogues we then apply in Sect.\ref{section_3} a tomographic analysis to compare the simulated results to the expected theoretical one and focus on the covariance matrix.\\
Appendices gives more technical details about some properties of the LN distribution, the Mehler transform and the tri-spectrum computation with it.
Throughout the paper we target a sub-percent precision of all our
spectra up to the Nyquist frequency.

%% file: section_0.tex
\section{The log-normal problem for filtered fields}
\label{sec:LN}

The autocorrelation of the matter (over-) density field is the Fourier
transform of its power-spectrum which, for an isotropic 3D field, reads 
\begin{align}
\label{eq:pktoxi}
 \xi(r)=\dfrac{1}{2\pi^2}\int_0^\infty dk ~k^2 P(k) \sin_c{kr},
\end{align}
where $\sin_c(x)=\dfrac{\sin x}{x}$ and the power-spectrum $P(k)$ can be computed with a
Boltzman solver as \texttt{CLASS}.
Technically such an integral can be computed efficiently with an
\texttt{FFTLog} algorithm \citep{Hamilton:2000}, which is the approach
we use in the following, or simply with an \texttt{FFT} by noticing
that $(r\xi(r),k P(k))$ form a Fourier (sine) pair.

\begin{figure}[htbp]
  \centering
  \includegraphics[width=\linewidth]{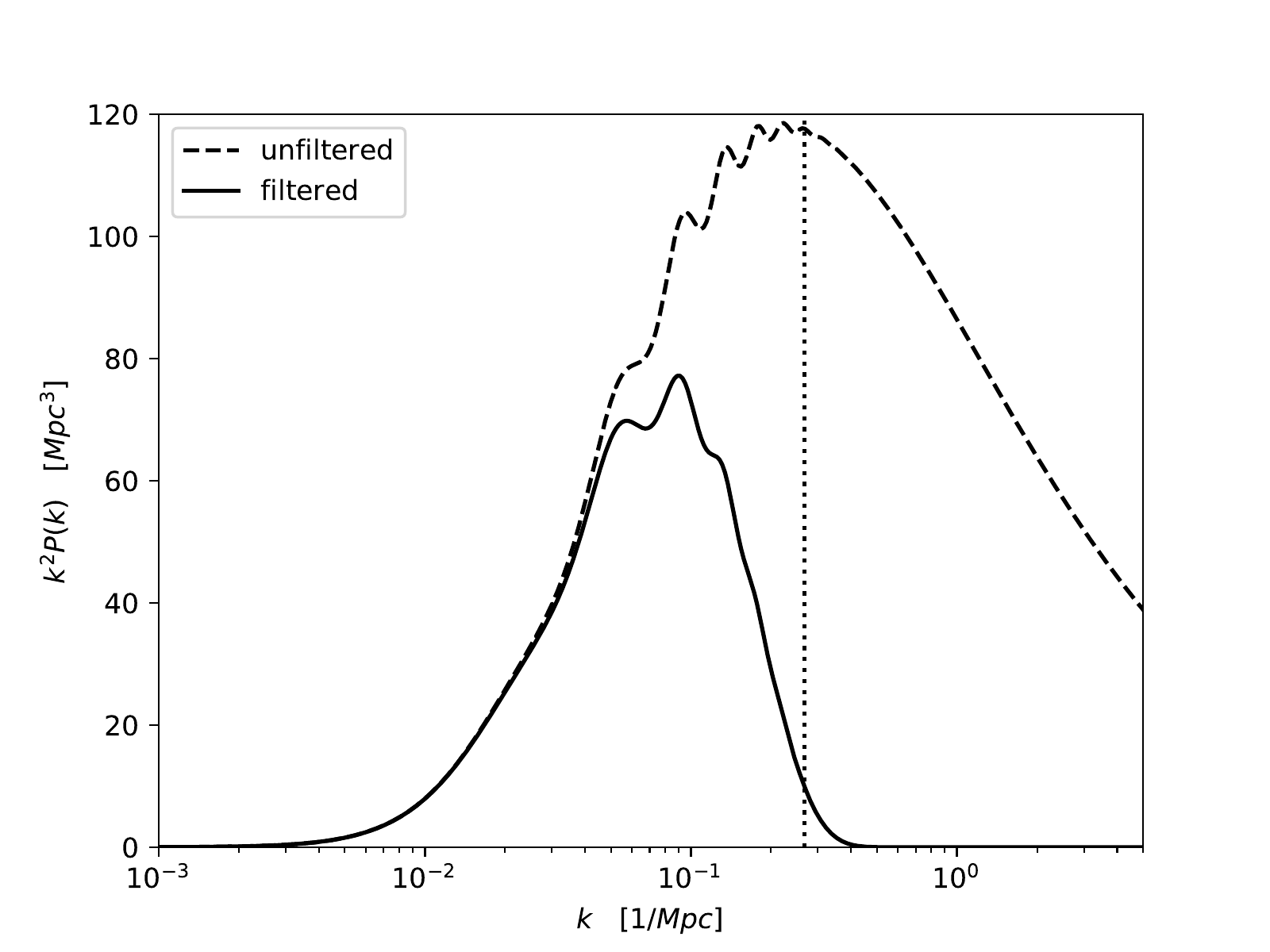}
\caption{\label{fig:filter} The linear power spectrum (dashed line) as computed by
    \texttt{CLASS} for a standard cosmology  and smoothed by a
   Gaussian window of radius $R_f=4\Mpc$ (full line). The vertical dotted line
    corresponds to $k_+=\frac{\pi}{2R_f}$ which is discussed in Sect \ref{sec:filtering}.}
\end{figure}

The variance of the field is by definition
\begin{align}
  \sigma^2=\xi(0)=\dfrac{1}{2\pi^2}\int_0^\infty dk ~k^2 P(k) 
\end{align}

but looking at the shape of the spectrum  (Fig. \ref{fig:filter}), one sees that the variance
will increase dramatically with the wavelength and actually even not converge
for a non-linear spectrum if no cutoff is introduced.
This is why fields are \textit{filtered} in
cosmology \citep[e.g][]{Coles:2003}. This, in practice, always happen
for a finite size experiment, but one may want to introduce explicitely
a smoothing window as a Gaussian one, modifying 
$P(k)\to P(k)e^{-k^2 R_F^2}$
which band-limits the spectrum below $k \lesssim \tfrac{2}{R_F}$.

Let us now recall some properties of a log-normal (LN) field, and
refer the reader to Appendix \ref{appendix_LN} for their demonstration.
Let $\delta^g(x)$ represent the Gaussian random field of the
density \textit{contrast}, then its log-normal transform in
cosmology is defined as 
\begin{align}
\label{eq:deltalog}
  \delta_{LN}(x)&= e^{\delta^g(x)-\tfrac{\sigma^2}{2}} -1,
\end{align}
where the different factors are here to ensure the mean to be zero.

Remarkably, the  Gaussian auto-correlation function $\xi_G(r)$ transforms as
\begin{align}
\label{eq:xitransform}
  \xi_{LN}(r)=e^{\xi^g(r)}-1.
\end{align}

This suggests a straightforward way to generate a LN field with some
target power-spectrum: just log-transform  (Eq. \ref{eq:deltalog}) a random Gaussian field
with a spectrum $P_\nu(k)$ corresponding to 

\begin{align}
\label{eq:lnxi}
\xi_\nu(r)=\ln \left(1+\xi_{LN}(r)\right).   
\end{align}
that we will call an \textit{inverse-log} transform.
From Eq.\ref{eq:xitransform} the LN field should then have the desired
spectrum.

\begin{figure}[htbp]
  \centering
  \includegraphics[width=\linewidth]{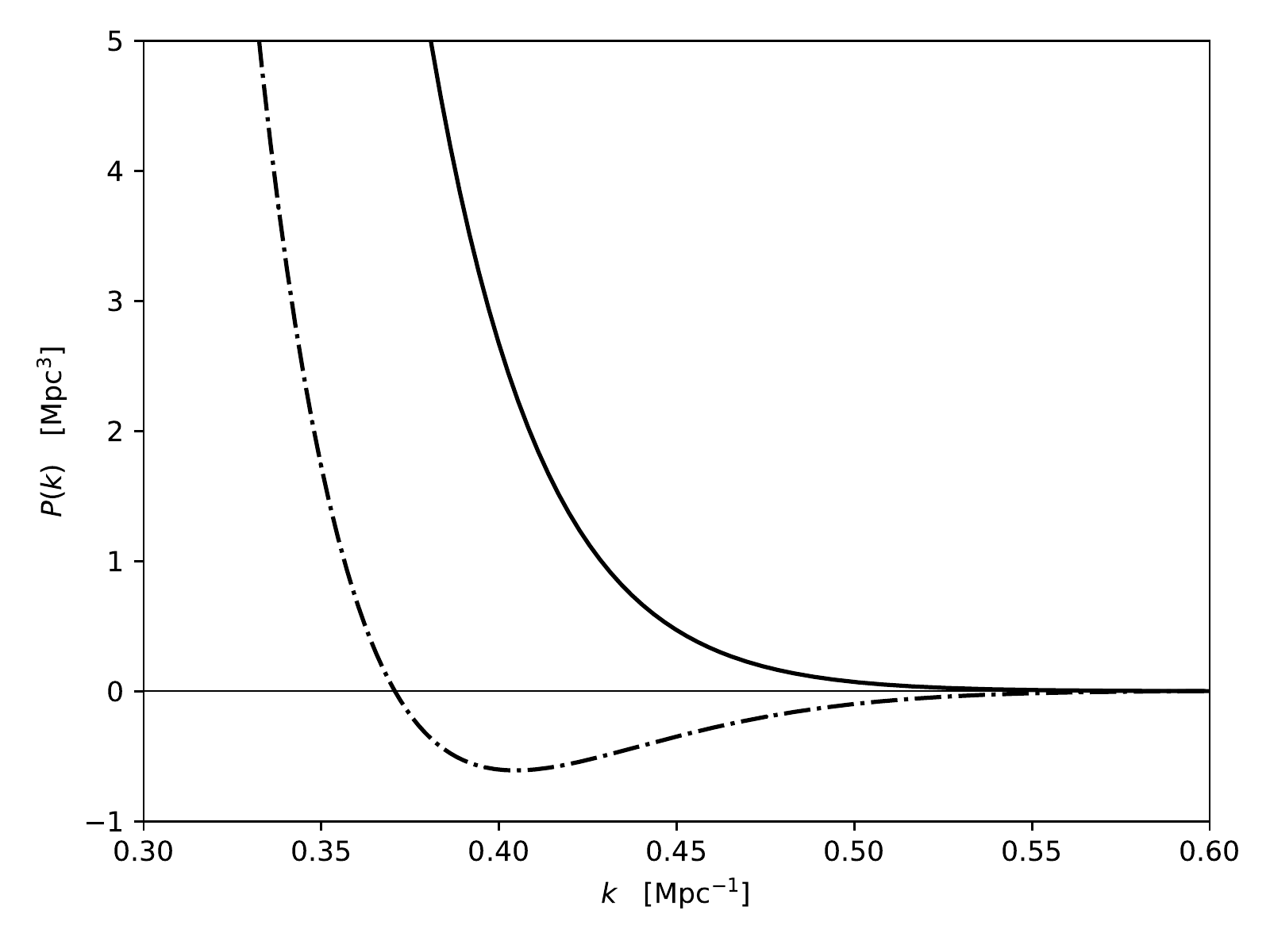}
  \caption{\label{fig:pkneg} The smoothed power
    spectrum (full line) and the one reconstructed by applying the
    Eq. \ref{eq:lnxi} inverse-log transform (dashed-dotted).
  }
\end{figure}

When performing that operation (Fig.\ref{fig:pkneg}) a problem
appears: at large $k$ the corresponding spectrum gets very small 
and becomes negative. 
This is clearly unphysical and one cannot use such a "spectrum" to generate any
Gaussian field.

A question one might ask is whether the very small negative values are
due to numerical issues (as machine precision or discretization, boundaries,
zero-padding in the FFT transform) or to more fundamental reasons. We
can get some insight about this question by using the following model.
For a smoothing radius of $R_f\simeq 4\Mpc$ the Gaussian window cuts the
cosmological spectrum in a region where it falls almost quadratically 
$P(k)\simeq \dfrac{A}{k^2}$ (with $A\simeq 100$ in our case). Then its filtered shape falls like

\begin{align}
\label{eq:pkmodel}
  k P(k)=\dfrac{A}{k}e^{-k^2 R_f^2}.
\end{align}
The associated auto-correlation (Fourier sine transform) is then
analytical. Indeed remembering that the $i/k$ operation corresponds to
integration in real space, or using \citet[][Eq. 3.954]{GradshteynRyzhik:2007}

\begin{align}
  r\xi_{LN}(r)&= \dfrac{1}{2\pi^2}\int_0^\infty k P(k) \sin(kr)~ dk \nn \\
&=\dfrac{1}{2\pi^2} \int_0^\infty \dfrac{A}{k}e^{-k^2 R_f^2}\sin(kr)~ dk \\
&=\dfrac{A}{4\pi}\erf\left(\dfrac{r}{2R_f}\right).
\end{align}
Expanding $\ln\left(1+\xi_{LN}(r)\right)$ in series \footnote{note that convergence of the series is not an issue because one can always normalize the Log-Normal field such that its variance is unity meaning that $|\xi_{LN}| \le 1$ }, we can compute the
spectrum of the corresponding Gaussian field as

\begin{align}
\label{eq:sum}
  kP_\nu(k)=&  4\pi \sum_{n=1}^\infty  \dfrac{(-1)^{n+1}}{n}\int_0^\infty
                \dfrac{ (r\xi_{LN}(r))^n}{r^{n-1}} \sin kr ~dr
\end{align}

The first term ($n=1$) corresponds to the Gaussian spectrum
(see Eq. \ref{eq:pkmodel} ). The higher order terms are smaller due
to the $\tfrac{1}{r^{n-1}}$ power suppression, unless the Gaussian term
gets very close to zero where they can predominate and lead to a
negative total contribution.
We have checked this by computing the integrals numerically,
keeping up to 30 terms in the sum to reach convergence, 
and verified that the values for $k\in[0.4,0.6]$ are indeed negative.

This completely independent method proves that the negativity of the
spectrum on Fig. \ref{fig:pkneg} is not a numerical artefact due
to some subbtle FFT effect.

So how can a "power-spectrum" be negative? The answer is actually simple:
\textit{any} function cannot represent an auto-correlation
function, it must be \textit{positive definite} \citep[for
an extensive discussion see][]{Yoglom:1986}. If we take the
autocorrelation values over a
discrete set of points, we end up from its very definition

\begin{align}
  \xi(r_{ij})=\xi_{i,j}=\E{\delta_i\delta_j}
\end{align}
to a \textit{covariance matrix} that must be positive definite.
The simplest way to check if the auto-correlation function is positive definite is to
study the sign of its Fourier transform. Since the inverse-log transform has a Fourier transform which is not guaranteed to be positive, 
it \textit{cannot} represent any genuine auto-correlation function.
This is very different from the direct log-transform (Eq.
\ref{eq:xitransform}) which is \textit{constructed} to
represent the auto-correlation of a transformed field.

We must conclude
that the very idea of constructing a LN field with a given
filtered spectrum, by transforming a Gaussian one, is mathematically ill-defined.

%% file: section_1.tex
\section{Sampling a field with a target \pdf and spectrum}
\label{sec:sampling}
Let us now consider the sampling of an isotropic field over a regular
cubic grid of step size
\begin{align}
a=L/N_s ,   
\end{align}
$N_s$ being the number of sampling points per dimension\footnote{From
  numerical considerations concerning FFT, a power of 2 is generally
  preferred} and $L$ the comoving box size fixed by the cosmology and the 
maximum wanted redshift.

Our goal is to obtain a proper power-spectrum up to the maximal
accessible frequency which is the Nyquist one 
\begin{align}
  k_N=\pi/a.
\end{align}

\subsection{Sampling a filtered field}
\label{sec:filtering}

We fist consider the case of a filtered field, as the solid line 
shown on Fig.\ref{fig:filter}.

\begin{figure}[htbp]
 \includegraphics[width=0.5\textwidth]{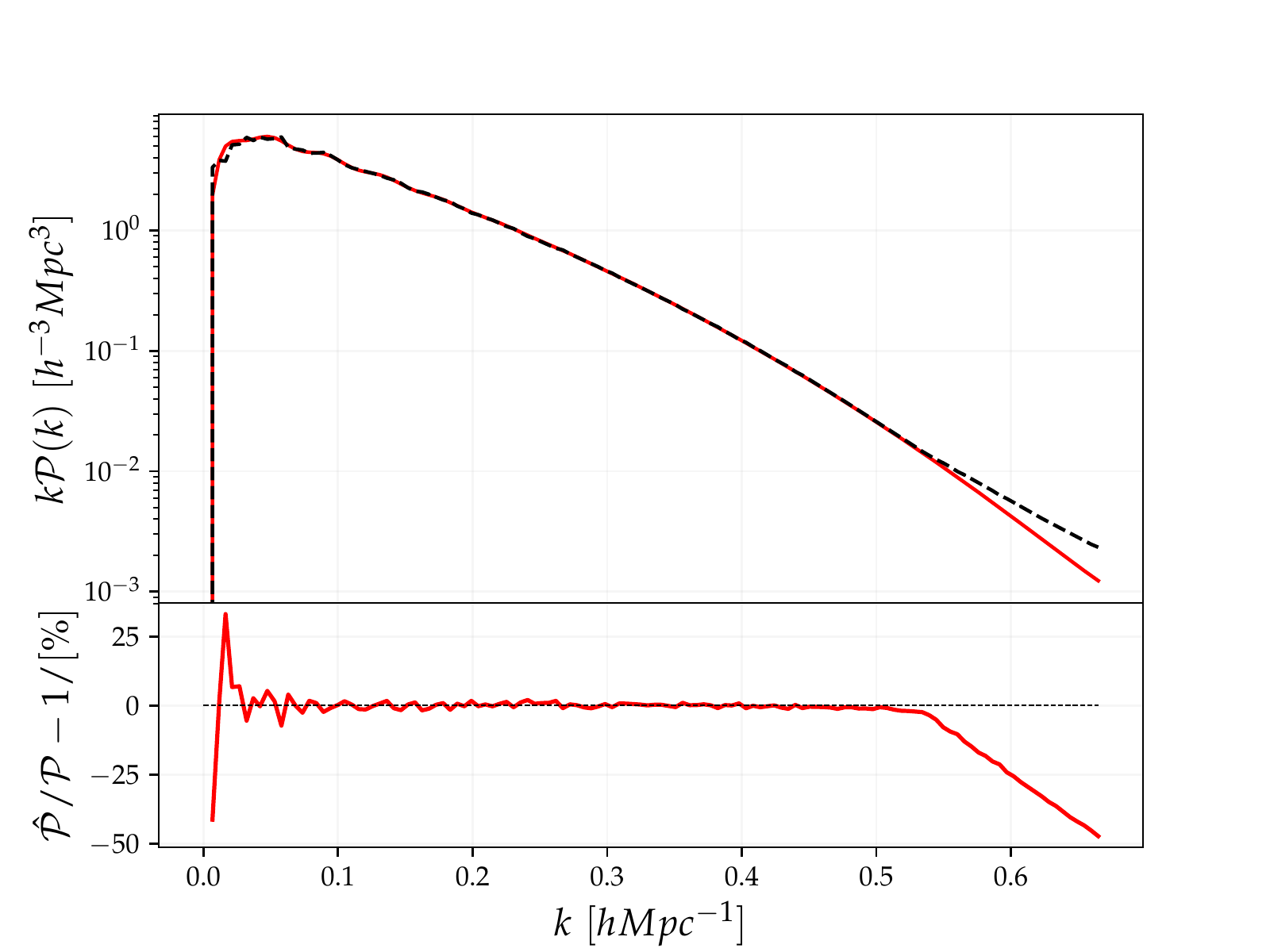}
 \caption{{\it Top: } Measured density power spectrum (solid red line)
   up to the Nyquist frequency in the case of a filtering scale of
   $R_f=5.5h^{-1}$Mpc with a sampling $N_{s}=256$ corresponding to a
   grid size $a = 4.7h^{-1}$Mpc, when the input spectrum negative
   values are clipped to 0.  The black dashed line is showing the
   expected power spectrum. {\it Bottom:} relative deviation between
   both.}
\label{fig:clipping}
\end{figure}

As we have seen in Sect.\ref{sec:LN} there is a mathematical problem
for the LN field when the spectrum becomes small leading to a negative
contribution to the required input one (Fig.\ref{fig:pkneg}). One may
think that the effect is so small that we can simply clip all negative
values to 0 to recover a valid spectrum.
Postponing the details of our full pipeline to Sect.
\ref{sec:pipeline}, we have use the clipped field to generate the LN one
and reconstructed the power spectrum that is compared to the target
one on Fig.\ref{fig:clipping}.
The result is unsatisfactory well before the Nyquist frequency.

\begin{figure}[htbp]
 \includegraphics[width=0.5\textwidth]{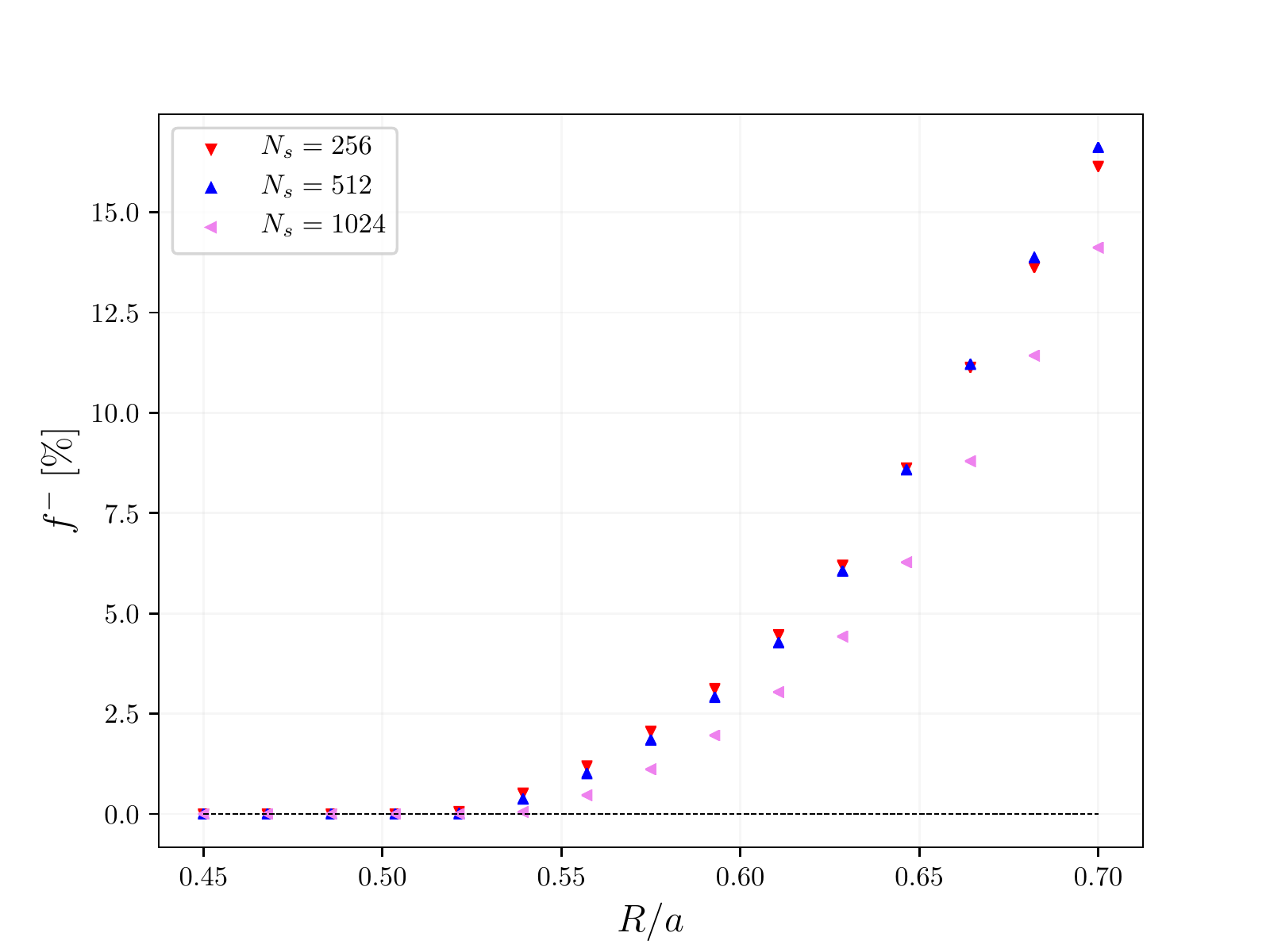}
 \caption{Fraction $f^-$ of negative values in the tridimensional
   $\mathcal P ^{virt} (\vec k)$ as a function of the relative
   filtering $R_f/a$ for three different grid samplings. The ratio
   $R_f/a$ represent the relative scale between the smoothing scale
   $R_f$ of the filtered power spectrum and the size of a grid unity
   $a$.} 
\label{negpower}
\end{figure}

Let us now try to see when the problem happens. To this purpose
we counted the fraction of modes with negative values with
respect to the smoothing size $R_f$ for 3 $N_s$ samplings (and fixed
box size,so $a$ values). 
The outcome of this test is
presented in Figure \ref{negpower}, which shows that in each case 
modes with negative power appear when $R_f \gtrsim a/2$.

Put it in another way, this means that we can reconstruct a proper LN field
as far as $R_f > a/2$.
This is only partially satisfactory since we may adjust the step size
(with $N_s$) to the smoothing radius, but we will only be able to
reconstruct the spectrum up to the Nyquist frequency
\begin{align}
\label{eq:kplus}
  k_+=\dfrac{\pi}{a}=\dfrac{\pi}{2R_F}.
\end{align}

As shown on Fig. \ref{fig:filter} as the dashed vertical line we will
not sample the spectrum entirely and miss some (tiny) power. This affects the
variance of the field although not much (about 2\% in this case).

There is no fully satisfactory solution since the procedure itself is
mathematically wrong whenever the spectrum reaches small values (the
$k_+$ cutoff avoids it). The full power LN sampling of a filtered field
cannot be achieved by transforming a Gaussian one. 

But why filter? In practice we use the simulation to generate a discrete set of 
\textit{galaxies} and the process introduces some filtering. If we
control \textit{exactly} this filtering
(what we will demonstrate in Sect.\ref{sec:catalog} we do not need to perform it
explicitly at the very start.
Then we can work
with an unfiltered field (as the dashed one on Fig.\ref{fig:filter})
that is well behaved for the transforms.
However as is clear from the figure there will always be some extra-power 
above the Nyquist frequency, so the key point is to handle properly \textit{aliasing}.

\subsection{The full pipeline}
\label{sec:pipeline}

We then start from an unfiltered field.
Although our method lies on a classical ground
\citep[e.g][]{Chiang:2013,Greiner:2015,Agrawal:2017}, we introduce two new
aspects:
\begin{enumerate}
\item we generalize the \pdf to any distribution,
\item we take into account aliasing to deal with the residual
  power.
\end{enumerate}

The idea to obtain any \pdf (for the density contrast $\delta$) is to go into
configuration space  and apply a non-linear local transform to the
Gaussian field ($\delta_\nu$)

\begin{align}
  \delta=\lcal(\delta_\nu).
\end{align}
The $\lcal$ function can be found easily by applying standard
probability transformation rules \citep{Bel:2016} and may need to be computed
numerically. In the LN case, the transformation is analytical and was given in
Eq.\ref{eq:deltalog}.

From now on, be $\delta(\vec x)$ a real, $L$ periodic and translational invariant field
with null expectation value, let us define  $\delta_\vec k$ as its
Fourier transform. On one hand, the translational invariance imposes
that the covariance between wave modes is diagonal,
i.e. $\langle\delta_{\vec k}\delta_{\vec k'}\rangle = \delta^{D}(\vec
k + \vec k')\pcal(\kvec)$, on the other hand the periodicity implies
that the Fourier transform $\delta_{\kvec}$ is non-zero only for
$\kvec = \vec n k_F$, where $k_F = 2\pi/L$ is the fundamental
frequency of the field and $\vec n$ is an integer vector. Adding the
fact that the field is real, it follows that the expectation value of
the square modulus of the Fourier transform is directly related to the power spectrum

\begin{equation}
\langle |\delta_{\vec{k}}|^2 \rangle=\frac{\mathcal P(\vec k)}{k_F^3} , 
\label{defpk}
\end{equation}
while the covariance between modes remains null. This property allows to set up a Gaussian field in Fourier
space by generating as two uncorrelated centered gaussian random
variables (the real and the imaginary part of the Fourier density
field $\delta_{\vec k}$). They must have the same variance which should be equal to half the
value of the power spectrum evaluated at the considered $k$-mode. This
is equivalent to generating the square of the modulus of $\delta_{\vec k}$ following an exponential distribution with parameter $\pcal(\vec k)/k_f^3$
and a random phase peaked from a uniform distribution between $0$ and $2\pi$.  
Thus, in practice the Fourier transform of a Gaussian field can be generated on a Fourier grid as

\begin{equation}
\delta_{\vec k} = \sqrt{-\mathcal P(\vec k)/k_f^3 \ln(1-\epsilon_1)} e^{2\pi \epsilon_2} \; , 
\label{field}
\end{equation}
with  $\epsilon_1$ and $\epsilon_2$ being two uncorrelated uniformly distributed between $0$ and $1$ random variables. 
In addition to the appealing property of having a null correlation
between different modes, generating a Gaussian field in Fourier space allows to take avantage of the Fast Fourier Transforms (3D FFT) algorithm.

The novel ingredient is to consider that since we are using here a "raw" (unfiltered) cosmological spectrum, more
power is leaking around the Nyquist frequency.

\begin{figure*}[h]
\begin{center}
\includegraphics[width=\linewidth]{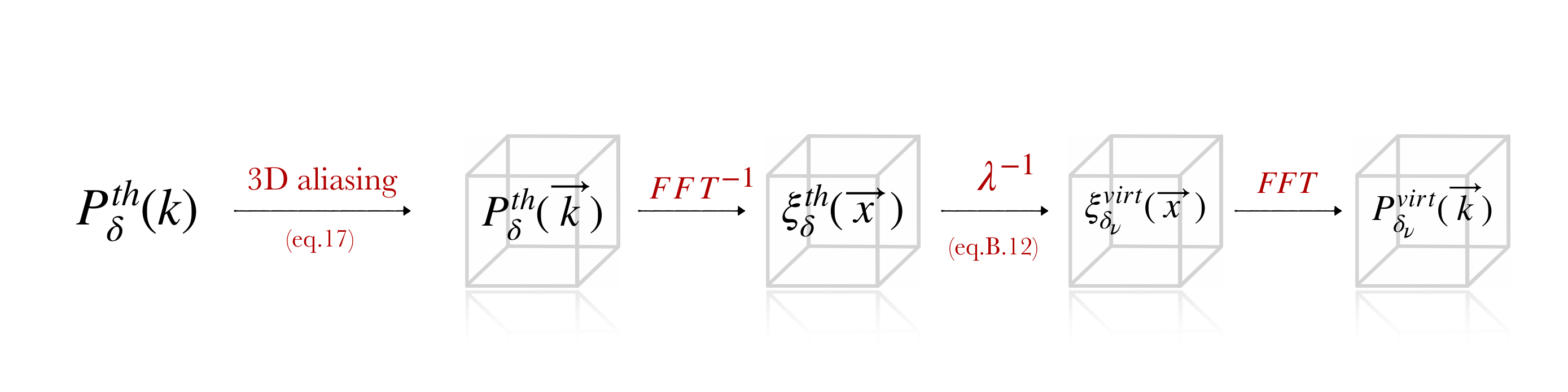}
\end{center}
\caption{Schematic view of the method used to build the virtual power
  spectrum $\mathcal P^{virt}_{\delta_\nu}(\vec k)$. The grey boxe
  symbol means that we consider $3$ dimensions. }
\label{schema}
\end{figure*}
\begin{figure}[h]
\centering
\includegraphics[width=.5\textwidth]{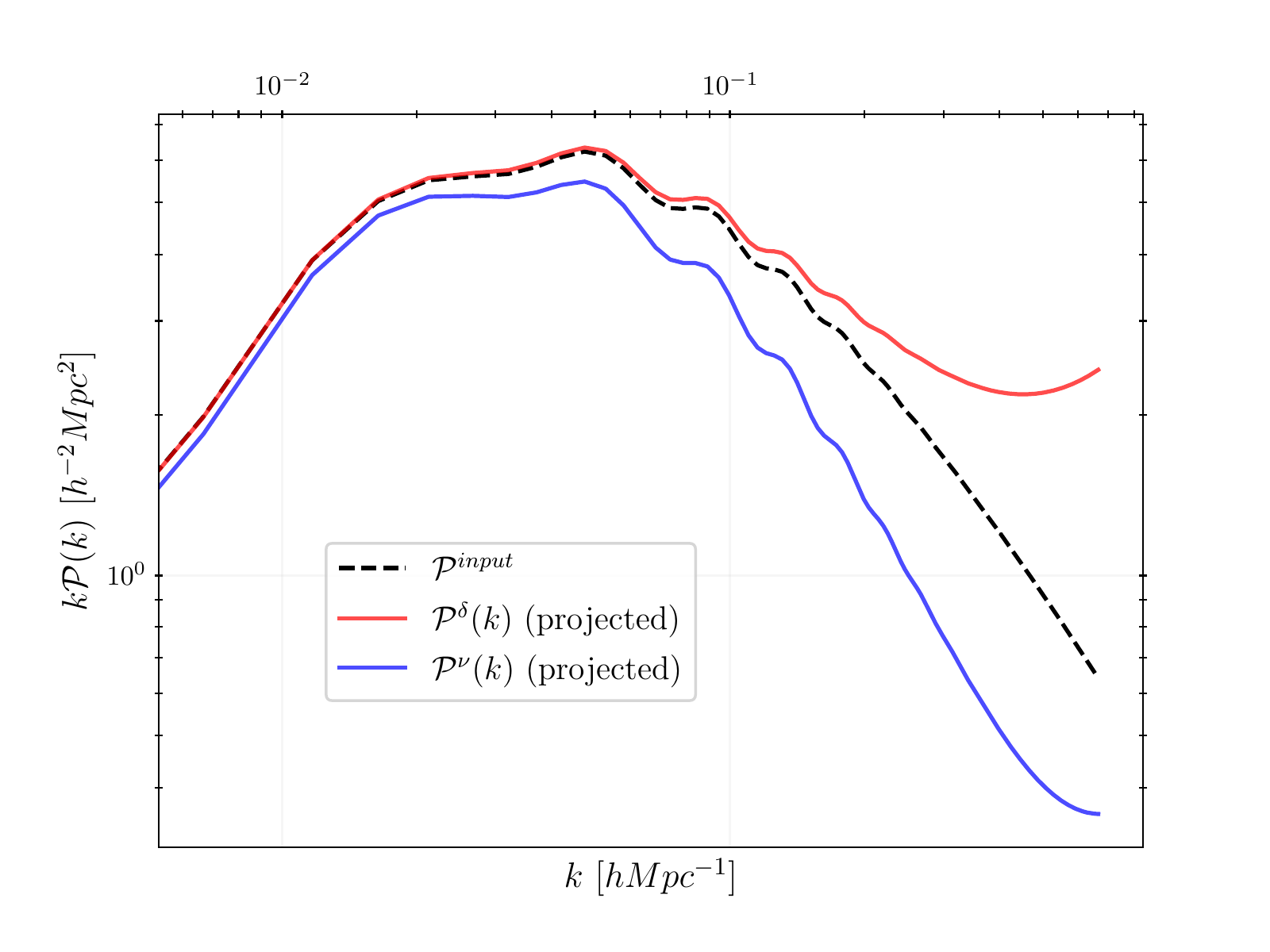}
\caption{Power spectra involved in the Monte Carlo process. In dashed
  black line is drawn the analytical one-dimensional matter power
  computed by \texttt{CLASS}. Then the shell-averaged power
  spectra (in shells of width $|\vec k|-k_f/2<|\vec k|<|\vec
  k|+k_f/2$) corresponding to the aliased version of the input power
  spectrum and the corresponding virtual power spectrum (see Fig.
  \ref{schema} for details) are respectively plotted in
  red and blue. All of them are plotted up to Nyquist frequency with a
  setting of $N_{s}=256$ and $L=1200h^{-1}$Mpc.}  
\label{fig1}
\end{figure}

Then to be coherent with our process of generating the Gaussian field with the required
input power spectrum we must take care of adding the aliased power
as an input in Fourier space \citep[see][for a detailed
review]{H&E88} which can be performed by summing the power spectrum aliases 

\begin{equation}
\hat{\mathcal P}(\vec k) = \sum_{\vec n}\mathcal P\left( |\vec k-2\vec nk_N| \right)\
\label{alias}
\end{equation}
where $\vec n$ is running over the 3D Fourier wavenumbers. Note that, since the aliasing effect is mixing modes which are uncorrelated the phases
remain uniformly distributed in Fourier space while the effective amplitude of the power spectrum is changing according to equation \ref{alias}.
In our analysis we have found that using only the first  $125$ contributions from $\vec n = (-2,-2,-2)$ to $\vec n = (2,2,2)$ is
enough to reach a percent level accuracy on the power spectrum of the
catalogue at the Nyquist Frequency. A more computationally
efficient choice would be to take only the first $27$ allias but it
would lead to an accuracy of around 5-6\%. In turn
if one decide to discard all alias contributions then nearly $2$\% of
the modes would be required to have a negative variance. Clipping
those pathological modes to 0 power, would
lead to a significant deviation of the power spectrum of the generated
field with respect to the expected one, even below the Nyquist
frequency. In the following will consider the first $125$ alias
contributions.



As detailed in \citet{Bel:2016} any local transform $\lcal$ applied to
a centered Gaussian field $\nu(\vec  x)$ corresponds to a one to one
mapping $\lambda$ of its $2$-point correlation function $\xi_\nu (\vec
r) \equiv \langle \nu(\vec x)\nu(\vec x + \vec r)\rangle$ such that
$\delta(\vec x) = \lcal[\nu(\vec x) ]$ and $\xi_\delta(\vec r) =
\lambda[ \xi_\nu (\vec r) ]$. The $\lambda$ function is given
explicitly in the Appendix (Eqs.\ref{eq:series} and \ref{eq:cn}).
As a result, using an inverse Fourier transform we can find the 3D
$2$-point correlation of the target non-Gaussian field $\delta(\vec
x)$, from wich, using the inverse mapping $\lambda^{-1}$ we are able
to compute the corresponding $2$-point correlation of the Gaussian
field $\nu(\vec x)$. In the end one only needs to Fourier transform
back in order to get the input power spectrum that will be indeed
characterising the input Gaussian field. In the following it will be
referred to as virtual: $\mathcal P^{virt}(\vec k)$ and one can notice
that being obtained from the Fourier transform of a regularly (grid)
sampled $2$-point correlation it already contains aliasing
effects. Thus, the input virtual Gaussian field can be generated on
the corresponding Fourier grid from equation \ref{field}. A summary of
the steps involved in the process computation of the virtual power
spectrum are shown in figure \ref{schema}.

Finally, we can inverse Fourier transform the realisation of the Gaussian field and apply to it a local transformation which will automatically turn both the \pdf and the power spectrum into the expected ones. It is clear that if the input power spectrum was not aliased (as it naturally is) then the corresponding inverse Fourier transform could not be interpreted as a regularly sampled Gaussian field, thus the process would not be self consistent.  
%
%
%

Figure \eqref{fig1} illustrates the power spectra involved in the
generation of the non-Gaussian density field. One can see the raw input power spectrum obtained from the \verb?CLASS? code and its corresponding aliased version ($\pcal^\delta(k)$). Note that the aliasing needs to be applied on the $3$-dimensional Fourier grid while we represent only the averaged power spectrum in each Fourier shell of size $k_f$. In addition one can see on the same figure the corresponding power spectrum of the Gaussian field that we use to generate Monte-Carlo realisations. We can notice an excess of power at large $k$ which corresponds to the aliasing contribution.

 \begin{figure}[htbp]
 \centering
 \includegraphics[width=0.5\textwidth]{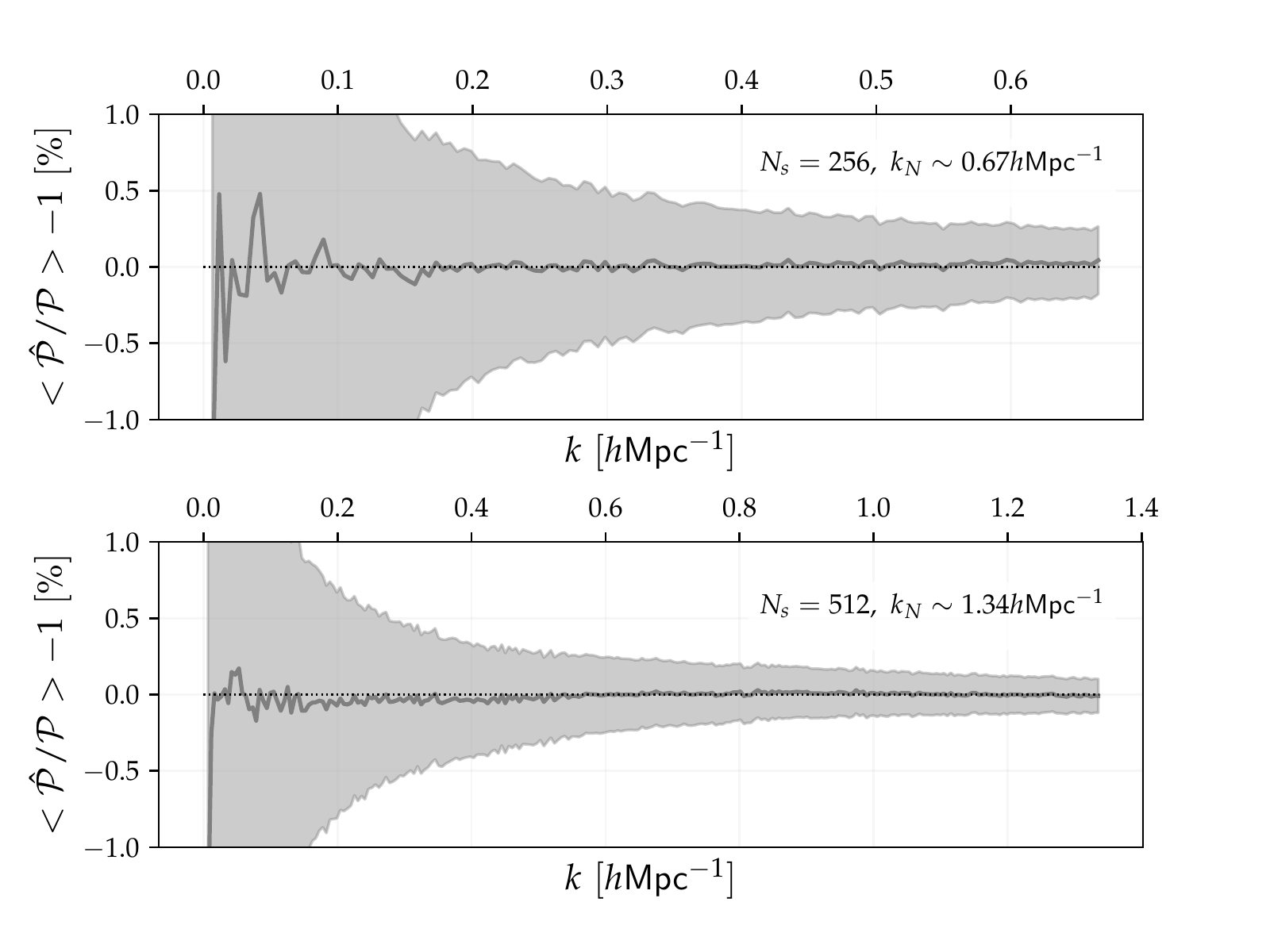}
 \caption{For 1000 realisations of the density field we compute the averaged 3D power spectrum that we compare relatively to the expected 3D power spectrum. We then compute the shell-averaged monopoles of this residuals in shells of width $|\vec k|-k_f/2<|\vec k|<|\vec k|+k_f/2$. The result is presented in percent with error bars. The used setting is a sampling number per side of 256 in the top panel and 512 for the other, all in a box of size $L= 1200 h^{-1}$Mpc at redshift $z=0$. Both results are computed up to the Nyquist frequency.}
 \label{modesmedian}
 \end{figure}

In order to verify the coherency of the method, we generate $1000$ realisations of Log-Normal non gaussian fields in a periodic box of size $L=1200 h^{-1}$Mpc with two different spatial resolutions corresponding to a number of sampling per side of  $N_s=256$ and $512$.  
%
%
From the definition of the power spectrum (Eq. \ref{defpk}), we
estimate the power spectrum on the $3$-dimensional Fourier grid by
computing the ensemble average of the $1000$ realisations and compare
it to the true expected power spectrum ($\pcal^\delta(\vec k)$). In
figure \ref{modesmedian}, we represent the $k$-shell averaged relative
difference between the estimated and expected power spectrum for each
individual wave modes. One can safely conclude that the accuracy of
the proposed method is better than $0.1$\% for wave modes close to the Nyquist frequency. In addition, no significant bias can be detected at the sub-percent level on the whole range of wave modes present in the density field independently from the choice of the spatial resolution.


\subsection{Covariance matrix}

A possible interest of being able to generate non-Gaussian fields with a Monte-Carlo method is related to generating a large number of realisations in order to estimate the covariance matrix (and its inverse) of a cosmological observable with a high level of statistical precision. In the following we define the shell averaged power spectrum as our observable and we estimate its covariance matrix between two shells respectively centered around wave numbers $k_i$ and $k_j$ as 

\begin{equation}
\hat C =\frac{\Delta^T\Delta}{N-1},
\label{covestim}
\end{equation}
where $\Delta$ is a matrix formed by the residual between the estimated power spectrum in each realisation and the estimated ensemble average of the power spectrum in each $k$-shell, $\Delta_{ij} = P_j(k_i) - \bar P(k_i)$ and $\bar P(k_i) = \frac{1}{N}\sum_{j=1}^NP_j(k_i)$, the $j$ index refers to the $j$-th realisation.  
If the deviation elements $\Delta_{ij}$ follow a Gaussian distribution
then one can show that the estimated covariance matrix elements
$C_{ij}$ follow a Wishart distribution. As a result, the estimator
\ref{covestim} is unbiased and the variance of the covariance matrix
elements \citep[see][]{Anderson} is given by 

\begin{equation}
\Var{C_{ij}} = (C_{ij}^2 + C_{ii}C_{jj})/(N-1).
\label{varc}
\end{equation}
 In the following we show that the statistical behaviour of the variance of the estimator of the power spectrum is in agreement with what we expect.

Having under-control both the target \pdf and the power spectrum we can predict to some extent the expected covariance matrix $C$ of our power spectrum estimator. Since the density field generated with a Monte-Carlo process is non-Gaussian, the covariance matrix of the estimator of the power spectrum involves contribution of the Fourier space $4$-point correlation function. For a translational invariant density field it reduces to

\begin{equation}
\langle \delta_{\vec k_1}\delta_{\vec k_2}\delta_{\vec k_3} \delta_{\vec k_4}\rangle_c =\delta^D(\vec k_1+\vec k_2 + \vec k_3 + \vec k_4)T(\vec k_1, \vec k_2, \vec k_3, \vec k_4)
\label{trispec}
\end{equation}
where $T$ is defined as the tri-spectrum which is the Fourier transform of the $4$-point correlation function in configuration space.
As shown by \citet{SZ&H99}, the covariance matrix elements of the power spectrum estimator can be expressed as 

\begin{equation}
C_{ij}=\frac{\mathcal P(k_i)^2}{M_{k_i}}\delta^D_{ij}+k_F^3\bar T(k_i,k_j)\ ,
\label{cij}
\end{equation}
where $M_{k_i}$ is the number of independent modes in shell $i$ and 

\begin{equation}
\bar T(k_i,k_j) = \int_{k_i}\int_{k_j}  T(\vec k_1, -\vec k_1, \vec k_2, -\vec k_2) \frac{\dif^3\vec k_1}{V_{k_i}}\frac{\dif^3\vec k_2}{V_{k_j}},
\end{equation}
where the integral is made over two shells of thickness $k_F$ centered and encapsulating respectively $k_i$ and $k_j$. The volume (in Fourier space) of each shell containing independent modes is denoted as $V_{k_i}$ and $V_{k_j}$, in the limit of thin shells we have that $V_k=2\pi k^2k_F$ thus $M_k = 2\pi k^2/k_f^2$. 
\begin{figure}[htbp]
 \centering
 \includegraphics[width=0.5\textwidth]{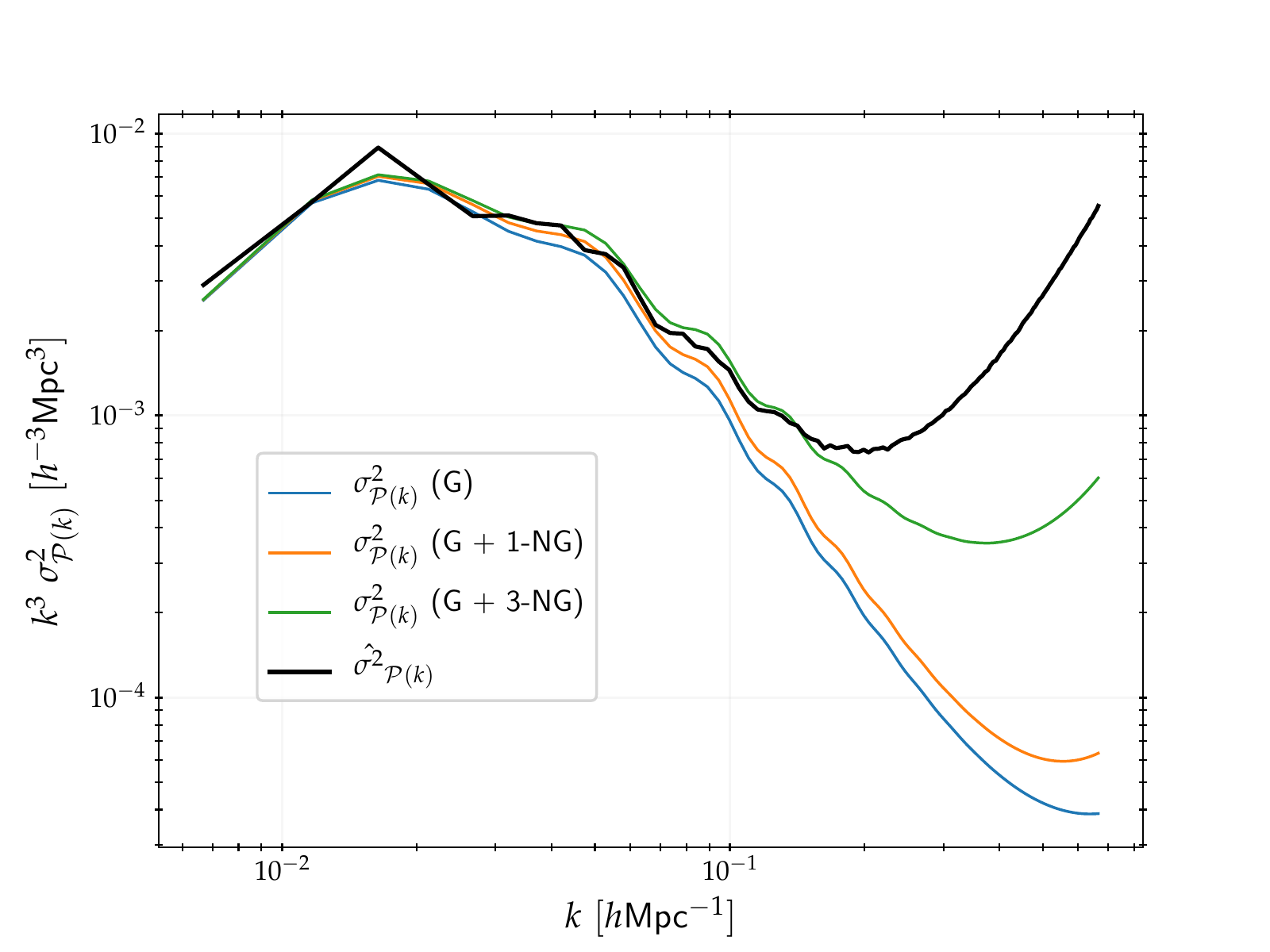}
 \caption{Measured diagonal of the covariance matrix for  7375 power spectra realisations of the density field using the Monte-Carlo method (black line). The other curves represent their predictions taking into account the gaussian part alone (G) or by adding some non gaussian contributions of equation \eqref{cij}. For example in (1-NG) one keeps only the term in $\mathcal P^3(k_i)$ in the trispectrum development presented in equation \eqref{tkiki} while in (3-NG) we keep all of them.}
 \label{figdiagcov}
 \end{figure}
 
 \begin{figure}[htbp]
 \centering
 \includegraphics[width=0.5\textwidth]{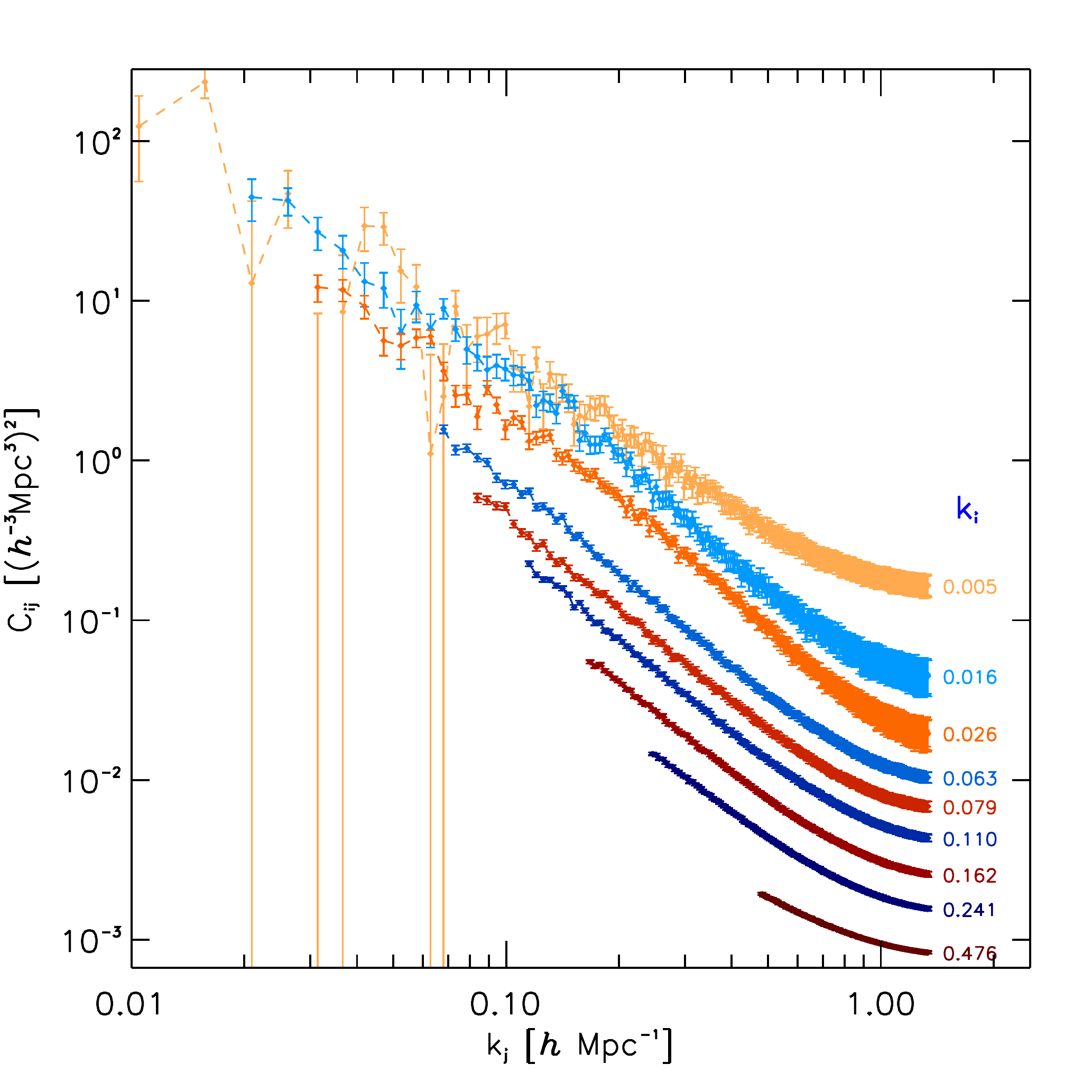}
 \caption{ Off diagonal elements of the covariance matrix estimated with $N=7375$ realisations, showing the dependance of the $C_{ij}$ with respect to $k_j$ at various fixed $k_i$ labeled on the right of the panel. The error bars are computed from equation \ref{varc}. }
 \label{offdiagCijPk}
 \end{figure}

In appendix \ref{appendix_trispectre}, we show how to predict in a perturbative way the tri-spectrum of the generated non-gaussian density field for any local transform. In particular, for the contribution of the tri-spectrum to the diagonal elements of the covariance matrix we obtain the expression 

\begin{multline}
\bar T(k_i, k_i) \sim  8 c_1^2 \left \{ 4c_2^2 + 3c_3c_1 \right \} \mathcal P^3(k_i) +\\ +24 \left \{ 3c_1^2c_3^2 + 4c_1c_2^2c_3+ 12c_1^2c_2c_4 \right \} \mathcal P^2(k_i)\mathcal P^{(2)}(k_i) + \\+144 c_1^2c_3^2\mathcal P^{(2)}(0)\mathcal P^2(k_i), 
\label{tkiki}
\end{multline}
where $P^{(2)}(k_i) \equiv \fcal[\xi^2] = \int \mathcal P(q) \mathcal P(|\vec q+\vec k_i|) \dif^3\vec q $ and the $c_n$ are the coefficients of the Hermite transform of the function $\lcal$ 

\begin{equation}
c_n = \frac{1}{n!}\int_{-\infty}^\infty \lcal(\nu) H_n(\nu)\frac{e^{\frac{-\nu^2}{2}}}{\sqrt{2\pi}}d\nu\; ,
\label{cns}
\end{equation}
where $H_n$ denotes the probabilistic Hermite polynomial of order $n$. From equation \ref{tkiki} one can see that the covariance matrix elements are expected to depend on both the chosen target power spectrum and the probability density distribution of density fluctuations.

We generate $7375$ realisations of a LN density field characterised by a $\Lambda$CDM power spectrum at redshift $z=0$, the $c_n$ are thus given analytically. We can evaluate the covariance matrix elements of the power spectrum estimator as a simple matrix product (see Eq. \ref{covestim}). In figure \ref{figdiagcov} we show the diagonal elements, namely the variance at each wave mode compared to the Gaussian contribution and the expected non-Gaussian contribution coming from equation \ref{tkiki}. It confirms that for intermediate wave modes the non-Gaussian correction starts being relevant. While it fails to reproduce the full $k$-dependance due to the fact that expression \ref{tkiki} has been obtained in a perturbative way. In Figure \ref{offdiagCijPk}  we show some combinations of modes $k_i$ and $k_j$ of the covariance matrix elements, they exhibit a clear dependance in such combination showing that due to the non-Gaussian nature of the created density field long and short wave modes are correlated in our power spectrum estimator.

In the following we will extend the case of the continuous sampled density field to the creation of a catalogue of discrete objects, which could be galaxies, clusters haloes or simply dark matter particles.

%% file: section_2.tex
\section{Production of a catalogue}
\label{sec:catalog}

\subsection{ Poisson sampling}

Simulating a galaxy catalogue implies transforming the sampled
continuous density field $\delta(\vec x)$ into a point-like
distribution. The density field must therefore be translated into a
number of objects (galaxies, haloes or dark matter particles) per cell
imposing an average number density $\rho_0$ in the comoving volume
such that $\rho(\vec x) = \rho_0[1+\delta(\vec x)]$ and performing a
Poisson sampling \citep{Layzer:1956}.  
To do so, one must choose an interpolation scheme in order to be able
to define a continuous density field $\rho^{(i)}(\vec x)$ between the
sampling nodes $\vec x_j$ surrounding  a cell centered on position
$\vec x_i$. This way for each cell $i$ one is able to compute the
expected number of object $\Lambda_i$ as  

\begin{equation}
\Lambda_i = \int_{v_i}\rho^{(i)}(\vec x) \dif^3\vec x, 
\label{lambi}
\end{equation}
where in practice the integration domain $v_i$ corresponds to the volume of a cell. 
Finally we assign to the cell the corresponding number of galaxies $N_i$ such  that the probability of observing $N$ objects given the value of the underlying field $\Lambda$ is given by a Poisson distribution $P_N =\frac{\Lambda^N}{N!}e^{-\Lambda}$. This way one can distribute the right number $N_i$ of objects in each cell volume with a spacial probability distribution function proportional to the interpolated density field $\rho^{(i)}(\vec x)$ within the cell.  


The most straightforward  interpolation scheme consists in populating
cells uniformly with the corresponding number of objects, which is called the Top-Hat scheme. 
One can guess that on scales comparable to the size of the randomly populated cells the power spectrum of the Poisson sample won't match the expected power spectrum.
Be $\tilde\delta (\vec x)$ the sampled density contrast field (the
true density contrast field multiplied by a Dirac comb) the
corresponding interpolated density contrast within the cell
$\hat\delta(\vec x)$ is obtained by convolving the sampled density
field with a window function $W(\vec x)$ leading in Fourier space to
the power spectrum relevant for the Poisson process as $\hat P(\vec k) =
\tilde P(\vec k) |W(\vec k)|^2$ where $\tilde P(\vec k)$ is the power
spectrum of the sampled density field, namely the aliased power
spectrum. One can thus finally obtain that the expected power spectrum
of the created catalogue will be  

\begin{equation}
\hat {\mathcal P}(\vec k) = |W(\vec k)|^2 \sum_{\vec n}\mathcal P\left( |\vec k-2\vec nk_N| \right) + \frac{1/(2\pi)^3}{\rho_0}, 
\label{poissonpk}
\end{equation}
where the additional term on the right corresponds to the shot noise contribution due to the auto-correlation of particles with themselves. Note that the Fourier transform of the chosen convolution function $W$ is cutting the power on small scale which is equivalent to smoothing the density field on the size of the cells. 

The interpolation scheme for the number density within each cell defines the form of the smoothing kernel $W$, in the following we consider two different interpolation scheme. The first one (the first order) is the Top-Hat which consists in defining cells around each node of the grid  and assigning the corresponding density within the cell. The second (the second order) is a natural extension which consists in defining a cell as the volume within $8$ grid nodes and adopting a tri-linear interpolation scheme between the nodes. In any of the two cases the window function \citep[see][for higher order smoothing functions]{SCSC16} takes the general form $W^{(n)}(\vec k) =\left [  j_0(k_xa/2)  j_0(k_ya/2)  j_0(k_za/2)\right ]^n$, where $j_0$ is the spherical Bessel function of order $0$ and the index $n$ corresponds to the order of the interpolation scheme. 
 
\begin{figure}[htbp]
 \centering
 \includegraphics[width=0.5\textwidth]{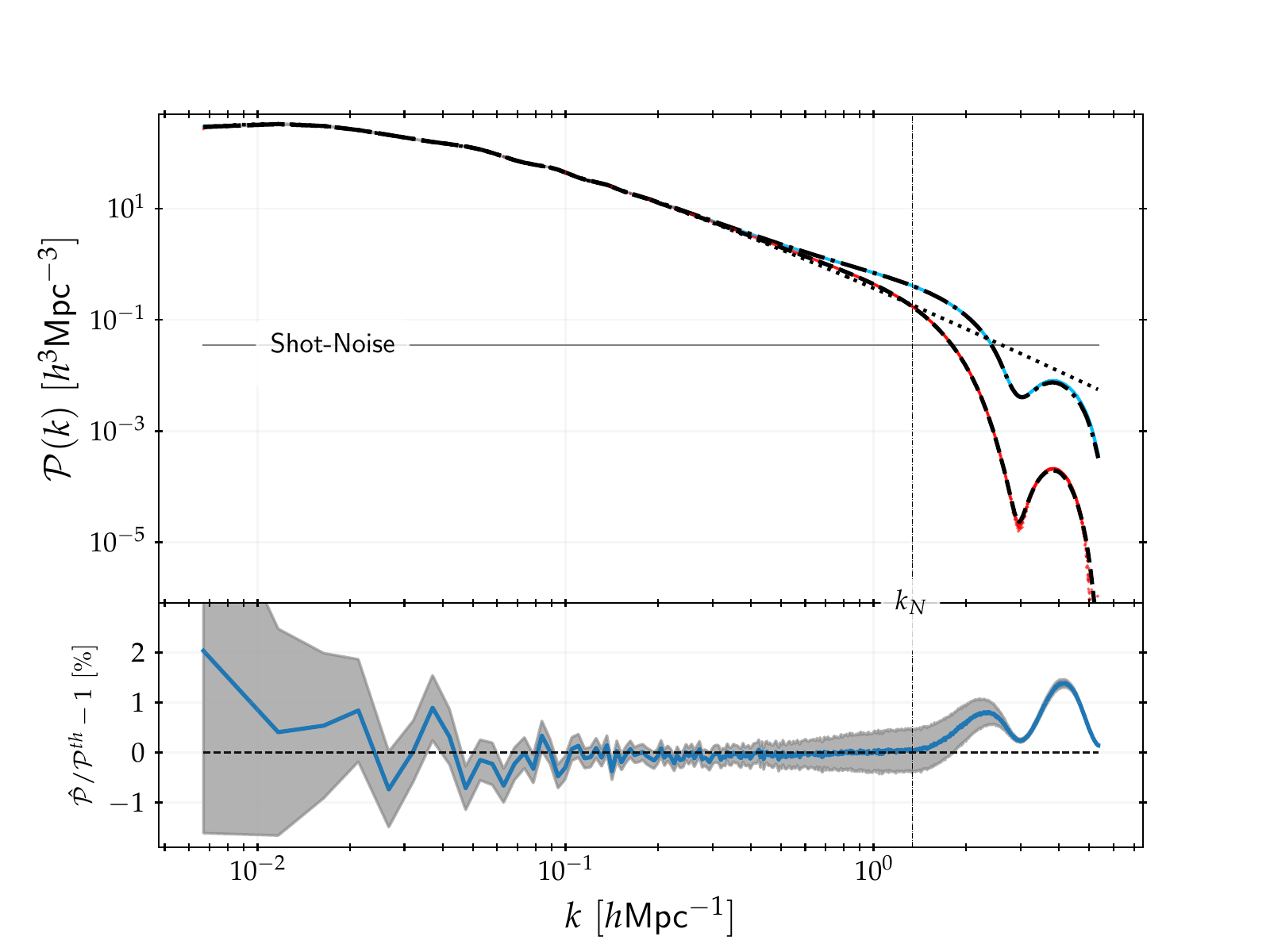}
 \caption{{\it Top: } Measured power spectra averaged over 100 realisations of the poissonnian LN field for the TopHat interpolation scheme (blue curve with prediction in semi-dotted black line) and for the linear interpolation scheme (red curve and prediction in dashed line). Note that the shot-noise is subtracted from measures (dotted horizontal line) and is about $3.48\times 10^{-2}\ h^3$Mpc$^3$. The dotted black curve represents the alias-free theoretical power spectrum computed by \texttt{CLASS}. {\it Bottom: } Relative deviation in percent between the averaged realisations (with shot-noise contribution) and prediction (with the same shot-noise added) in blue line with error bar in grey for the TopHat interpolations schemes. Snapshots are computed for a grid of size $L=1200\ h^{-1}$Mpc and parameter $N_s=512$. Here comparisons are made well beyond the Nyquist (vertical line) frequency at $k_N \sim 1.34\ h$Mpc$^{-1}$.}
 \label{poissonpkfig}
 \end{figure}

We estimate the power spectrum of the catalogues with the method
described by \citet{SCSC16} employing a particle assignment scheme of
order four (Piecewise Cubic Spline) and the interlacing technic to
reduce aliasing effects. Note that these choices are intrinsic to the
way we estimate the power spectrum of the distribution of generated
objects and has nothing to do with the way we generate the catalogues.
In figure \ref{poissonpkfig}, we compare the power spectra of the
catalogues of objects in case of the two interpolation schemes
described above.  We see that as expected the linear interpolation
scheme reduces more the extra power (due to aliasing) on small scales.  

In the same figure we also show the expected power spectra computed
with equation \ref{poissonpk} and corresponding to the two mentioned
interpolation schemes.
We demonstrate in both cases that we control precisely the smoothing of
the spectrum up to the Nyquist frequency and even above.

\subsection{Light cone}

In the following we describe how we build a light cone from our
catalogue  and compare two methods.

{\it Shell method:} The first idea is to glue a series of comoving
volume at constant time in order to reconstruct the past light cone
shell by shell \citep{mice1,mice2}. We first select a redshift interval $\Delta z$ labeled by $z_{{\rm min}}$ and $z_{{\rm max}}$ and a number $N_{{\rm shl}}$ of shells within it.  
For each of these shell, we generate a point-like distribution in
a comoving volume at constant cosmic time. Obviously, we perform the
poisson sampling only of the cells contributing to the considered
redshift shell. In addition, we keep only the objects belonging to the
comoving volume spanned by the redshift shell, defined by $[R(z_i -
dz/2),R(z_i + dz/2)]$ where $z_i$ corresponds to the redshift of the
comoving volume, $dz=\Delta z/N_{shl}$ and $R(z)$ is the radial
comoving distance.   
The light cone covers $4\pi$ steradians of the sky. 
In the next section we will show the effect of the choice of the
number of shells $N_{{\rm shl}}$ used to build the light cone,  on the
angular power spectrum. 

{\it Cell method:} The second method is faster. Rather than simulating many redshift-shells, one
selects a single redshift $z_0$ chosen at the middle of the radial
comoving space spanned by the light cone. 
We generate the corresponding Gaussian field in a comoving
volume on a grid at $z=z_0$. At this level one needs to include some
evolution in the radial direction from the point of view of an
observer located at the center of the box.  To do so, there are two
possibilities. 
\begin{itemize}
\item One can simply think of rescaling the Gaussian field at
a comoving radial distance $x(z)$ (from the observer) with the
corresponding growth factor $D(z)$ which rules the evolution
\citep{Peebles80} of linear matter perturbations. In this way it is clear
that as on large scale the power spectrum of the density field will
follow the expected evolution in $D^2(z)$, however the small scales
will be affected in a non trivial way leading to a modification of the
shape of the power spectrum, 
\item or one can change the contrast field so that the
evolution of the density field will follow the growth factor $D(z)$.
For the LN case this would read
  \begin{align}
  \delta_{LN}(x,z)&= e^{\delta^g(x,z)-D^2(z)\tfrac{\sigma^2(0)}{2}} -1.
\end{align}

\end{itemize}
The second option is particularly well suited when generating a
density field following a linear evolution.
However the first one, although not exact, can allow for the fast
computation of spectra evolution for more complex cases as when
$D(k,z)$.
In the following section we compare, in the linear regime, the
shell-method and the two cell-methods in the case of the Log-Normal
density field.

%% file: section_3.tex
\section{Application to tomography}
\label{section_3}

In cosmology, several arguments can be put forward to justify a
tomographic approach. Unlike the estimation of the power spectrum or
the $2$-point correlation function, no fiducial cosmology needs to be
assumed in order to estimate the observable \citep[see][]{B&D11,
  M&D12, Asorey12}. Only angular observed positions and measured
redshift are required, thus making it a true observable quantity. In
addition, the observable is defined on a sphere simplifying its
combinations with other cosmological probes such as lensing
\citep{C&B12, Gaztanaga12} or CMB and H$\alpha$ intensity mapping.  

\subsection{Angular power spectrum $C_\ell$}



So far we worked on a Fourier basis but it is usefull to expand the
matter perturbations into spherical harmonics \citep{Peebles80} and consider its
coefficients

\begin{equation}
\delta_\ell^m(r) = \int_S \delta(r,\theta,\phi){Y_\ell^m} ^\ast(\theta,\phi)d^2\Omega\ , 
\end{equation}
%
Assuming that the field is statistically invariant by rotation, i.e.
its angular $2$-point correlation function only depends on the angular
separation and not on the absolute angular position (the analogue of
translational invariance in 3D) on the sky,  the
two point correlation of the harmonic coefficients depends only on the
order $\ell$, thus $C_\ell(r,r')\equiv \left<
  \delta_\ell^m(r){\delta_\ell^m}^\star(r') \right>$ is defined as the
angular power spectrum between shells $r$ and $r'$. 
We may relate this spectrum to the isotropic 3D one as

\begin{equation}
	C_\ell(r,r')=(4\pi)^2\int_0^\infty k^2 \mathcal P(k) j_\ell(kr)j_\ell(kr')dk
	\label{defcl}
\end{equation}
where $j_\ell$ is the spherical Bessel function of order $\ell$.
However, in this expression there is an explicit dependence
on the radial comoving distances $r$ and $r'$. 

One can project
the density field of a \textit{thick} redshift shell with a given weighting
function $W(z)$ and define our observable density field as
\begin{align}
\tilde\delta_1(\theta, \phi) = \int W(z)\delta(r(z), \theta,
\phi)\dif z.
\end{align}

The corresponding angular power spectrum \footnote{in the following we
  only consider auto-correlations between shells} can  be
predicted from 

\begin{align}
  \begin{split}
	C_\ell(z_1,z_2)=&(4\pi)^2\int dz dz^\prime W(z)W(z') \\
    & \quad \cdot \int_0^\infty k^2 \mathcal P(k) j_\ell(kr(z))j_\ell(kr'(z))dk.    
  \end{split}
	\label{clz}
\end{align}
 In practice, the numerical evaluation of equation \ref{clz} is not
 simple and we will use the
\texttt{Angpow} software \citep{Angpow:2018} which is fully optimised for this task.

This theoretical quantity may then be compared to our simulations by
considering the number counts within pixels from samples in the
$z_1<z<z_2$ range (i.e using for $W$ a top-hat window) since in our case the
only source of fluctuations is due to the over-density field.
We then simply project the objects of the catalogue on the sky, count them, normalize by the mean value within
pixels $\bar N$,
and compute the spherical power-spectrum $\hat C_\ell$ with the
\texttt{Healpix} \citep{Healpix:2005} software using the parameter $\text{nside}=2^{11}$. The shot-noise
contribution is classically $\tfrac{1}{\bar N}$. Note that as for the figure (\ref{poissonpkfig}), one compute the angular power spectrum at scales smaller than the grid pitch. In a spherical basis, the equivalent of the Nyquist mode is obtained using $\ell_N \sim R[z_{\text{mean}}]k_N$ where $z_{\text{mean}}$ is the averaged redshift of the particules composing the catalogue.

\begin{figure}[htbp]
\centering
\includegraphics[width=0.5\textwidth]{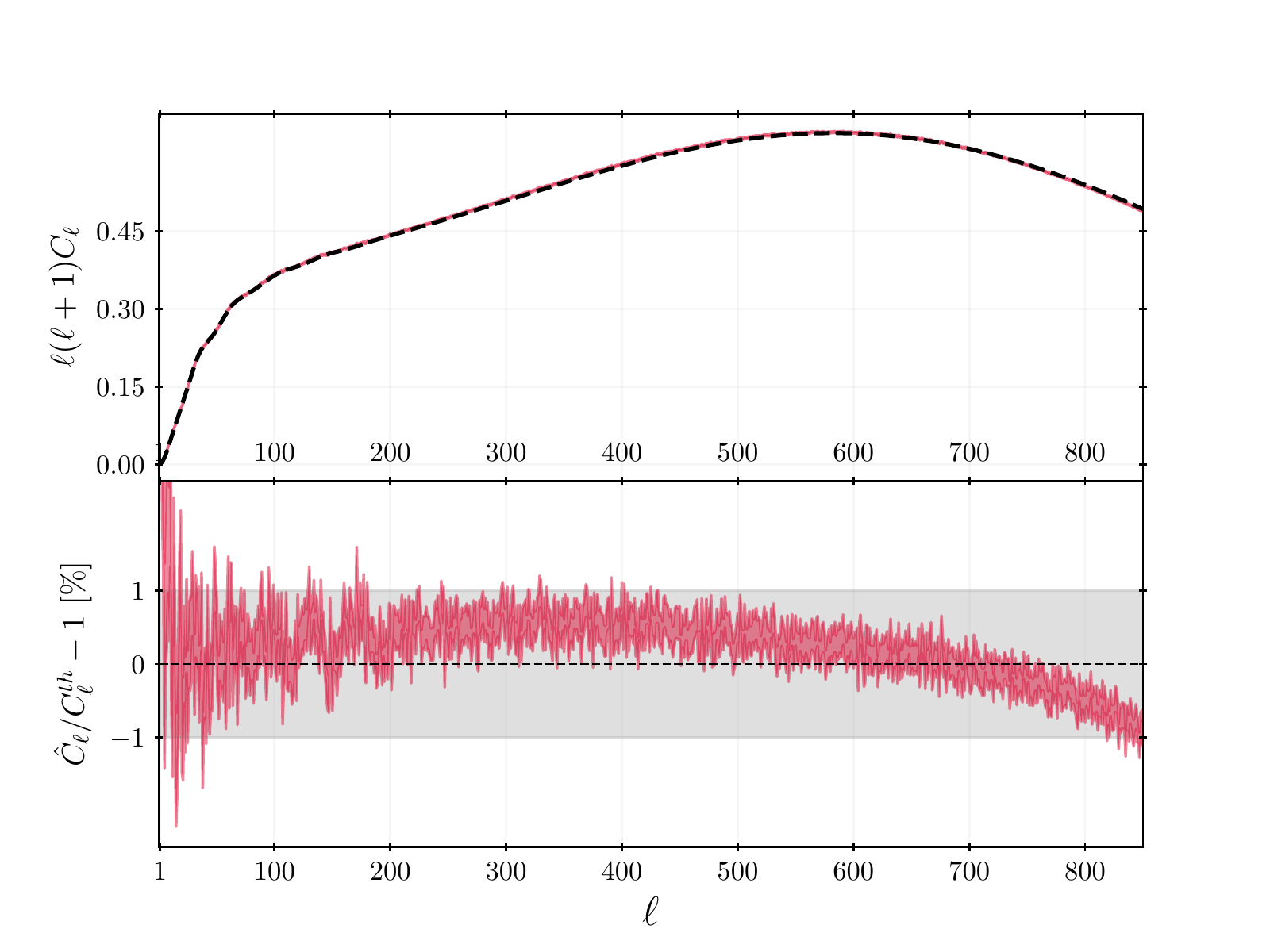}
\caption{{\it Top panel:} Thousand averaged $C_\ell$'s for simulated light cones using the shell-method with error bars (red curve) and corresponding prediction (dashed black curve). We simulate here a lightcone between redshifts $0.2$ and $0.3$ in a sampling $N_s=512$ and a number of shells $N_{shl}=250$ to ensure a sufficient level of continuity in the density field. The spherical Nyquist mode is situated around $\ell_N\sim 650$. {\it Bottom panel:} relative deviation in percent of the averaged $C_\ell$'s from prediction with error bars in red.}
\label{clstata}
\end{figure}

In Figure \ref{clstata} we compare the estimated angular power
spectrum in the $[0.2,0.3]$ redshift range to the predicted one (Eq. \ref{clz}) with the shell
method described in section \ref{sec:catalog} ($N_{shl} = 250)$ for thousand generated light-cones. In
the lower panel of the same figure we display the relative difference
between the two showing that the agreement is better than the percent
level.

\begin{figure}[htbp]
\centering
\includegraphics[width=0.5\textwidth]{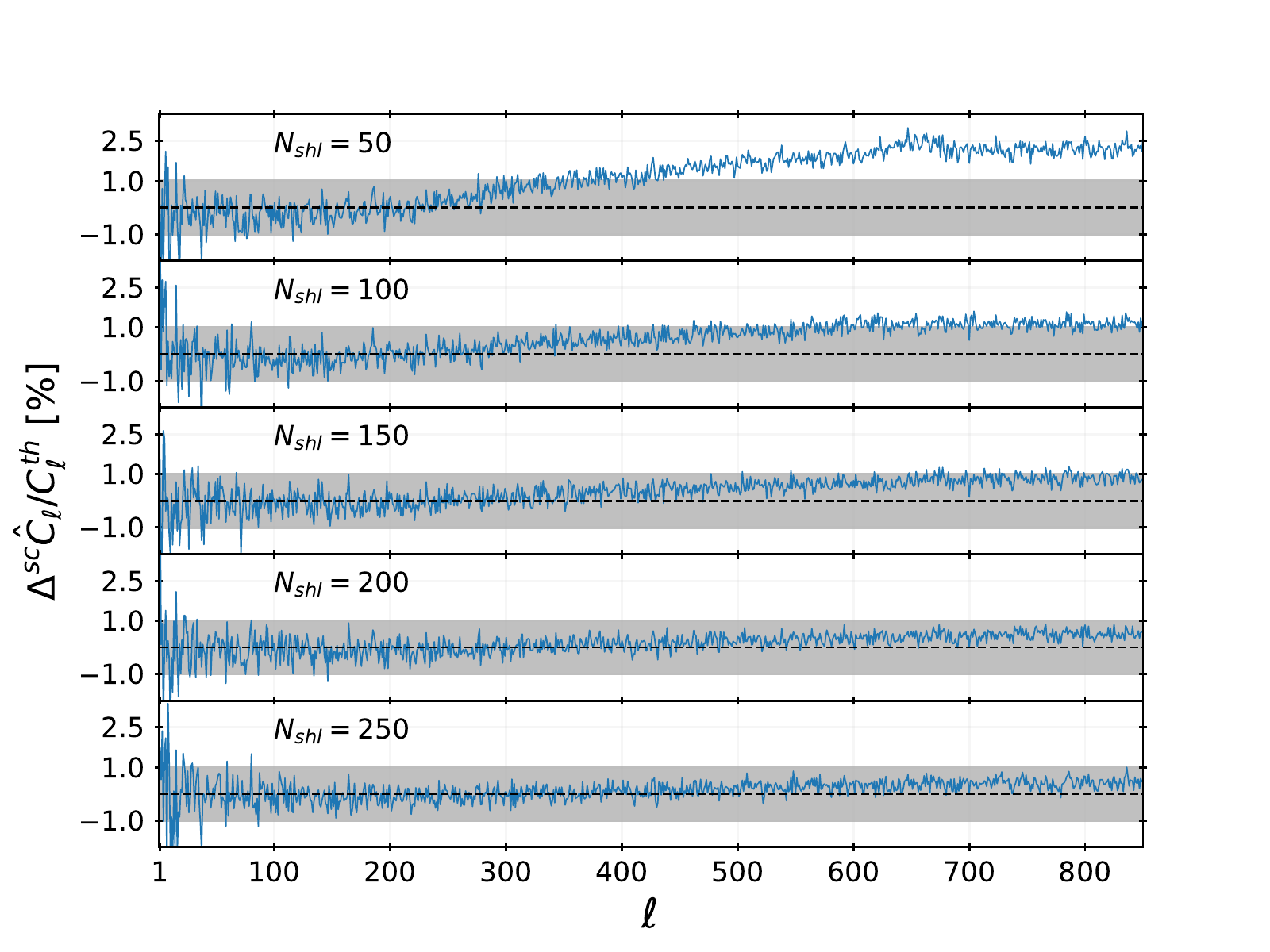}
\caption{Relative difference in percent between shell-method and cell-method for varying number of shells}
\label{clstatb}
\end{figure}

In order to quantify the impact of the choice of the number of shells in the shell method, we run a test comparison between the cell method and various number of shells $N_{shl}$ in the cell method. Note that since we are using a power spectrum which is evolving linearly across cosmic time we know that the shell method is expected to converge to cell method if we rescale the density field with the linear growing mode $D(z)$ as described in section \ref{sec:catalog}. Figure \ref{clstatb} shows the outcome of this analysis, which shows that in the considered redshift range the shell method is indeed converging to the cell method below the percent level as long as the number of shell is higher than $200$. 

\begin{figure}[htbp]
\centering
\includegraphics[width=0.5\textwidth]{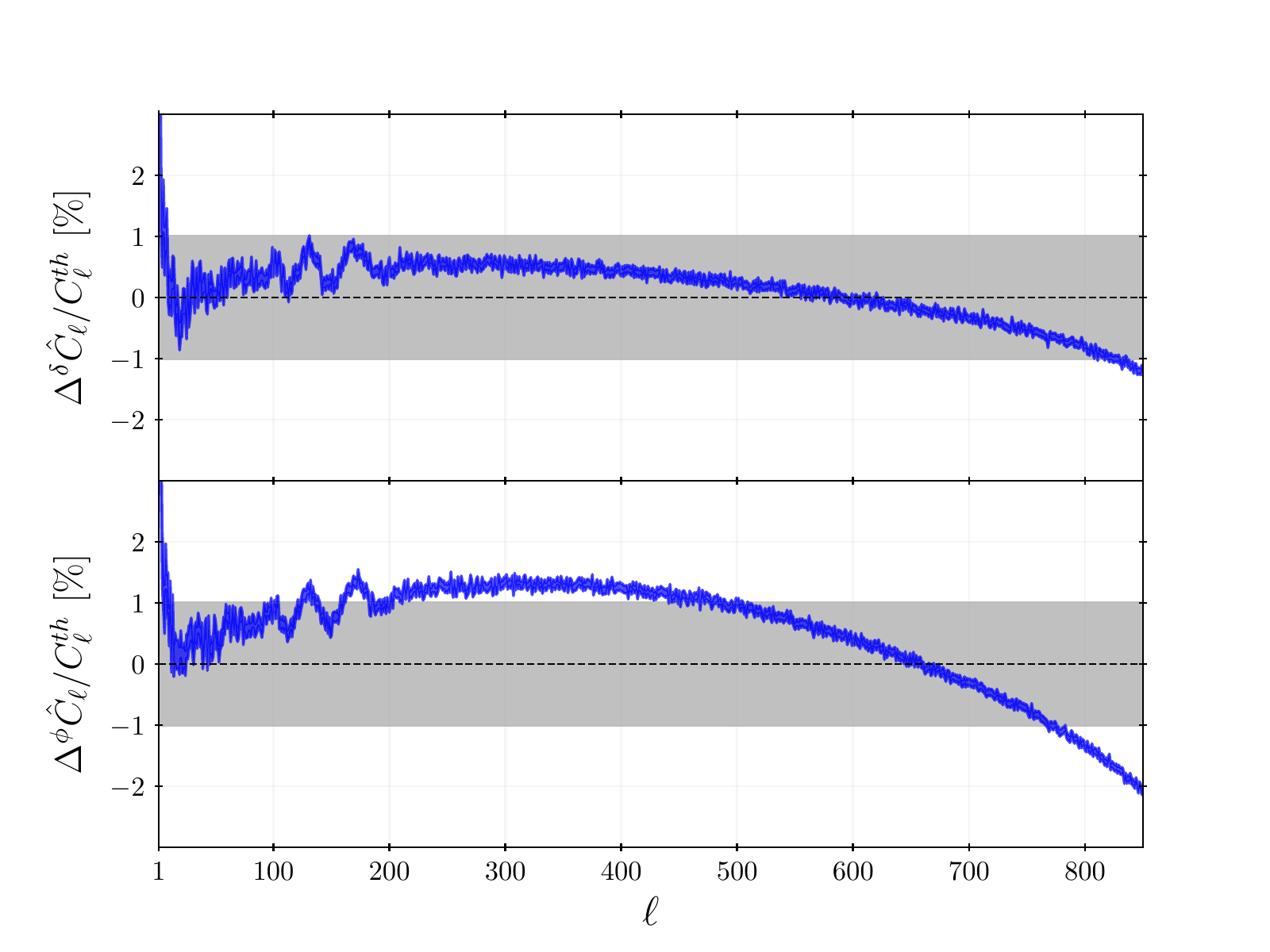}
\caption{Relative deviation in percent with error bar for 10000 averaged realisations of $C_\ell$'s in the context of the cell-method. In the top panel, the density field (non gaussian) is rescaled using linear growth function while in the bottom panel, the gaussian field following the virtual power spectrum is rescaled. The spherical Nyquist mode is situated around $\ell_N\sim 650$.}
\label{clstatc}
\end{figure}

Finally we make a comparison within the cell method, rescaling the
Gaussian field instead of rescaling the density field. This way we can
quantify the deviation when assuming that the Gaussian field evolves
linearly when instead it is the density field which is evolving
linearly. In Figure \ref{clstatc}, we show the relative deviation in
the two cases with respect to the expected power spectrum. One can see
that the deviation, despite being systematic, remains small (around
the percent level). Therefore considering these two cell-methods and as stated in the previous section, only the one offering better results will be recommended: the rescaling of the density field (top panel).


\subsection{Covariance matrix}

In this section we consider the cell method with linear rescaling of
the density field. We aim at estimating the covariance matrix of the
$\hat C_{\ell}$ estimator defined as 

 \begin{equation}
 \hat C_\ell \equiv \frac{1}{2\ell + 1} \sum_{m=-\ell}^\ell \tilde\delta_{1,\ell}^m \tilde\delta_{1,\ell}^{m \star},
 \label{clestimb}
 \end{equation}
with a high level of precision. Let us first show that the covariance matrix has a similar structure as the one of the power spectrum estimator studied in section \ref{sec:sampling}. By definition the covariance of $\hat C_{\ell}$  is 

\begin{equation}
	C_{\ell\ell'} \equiv \left< \hat C_\ell \hat C_{\ell'}\right>-\left< \hat C_\ell\right>\left< \hat C_{\ell'}\right>, 
	\label{cllp}
\end{equation}
where we can substitue $\hat C_\ell$ with its expression (see Eq. \ref{clestimb}). One immediately see that the first term of equation \ref{cllp} will let appear a $4$-point moment which can be expanded \citep{Fry84a,Fry84b} in terms of cumulent moments (or connected expectation values). It follows that it takes the general form

\begin{equation}
	C_{\ell\ell'} = \frac{2C^2_\ell}{(2\ell+1)} \delta^K_{\ell \ell'} + \bar T_{\ell\ell'}, 
\label{covcls}
\end{equation}
%
where $\bar T_{\ell\ell'}$ accounts for non-Gaussian contribution. Instead, in case of $\tilde\delta_{\ell}^m$ being a Gaussian field one can see that the covariance matrix is diagonal. 

\begin{figure}[htbp]
\centering
\includegraphics[width=0.5\textwidth]{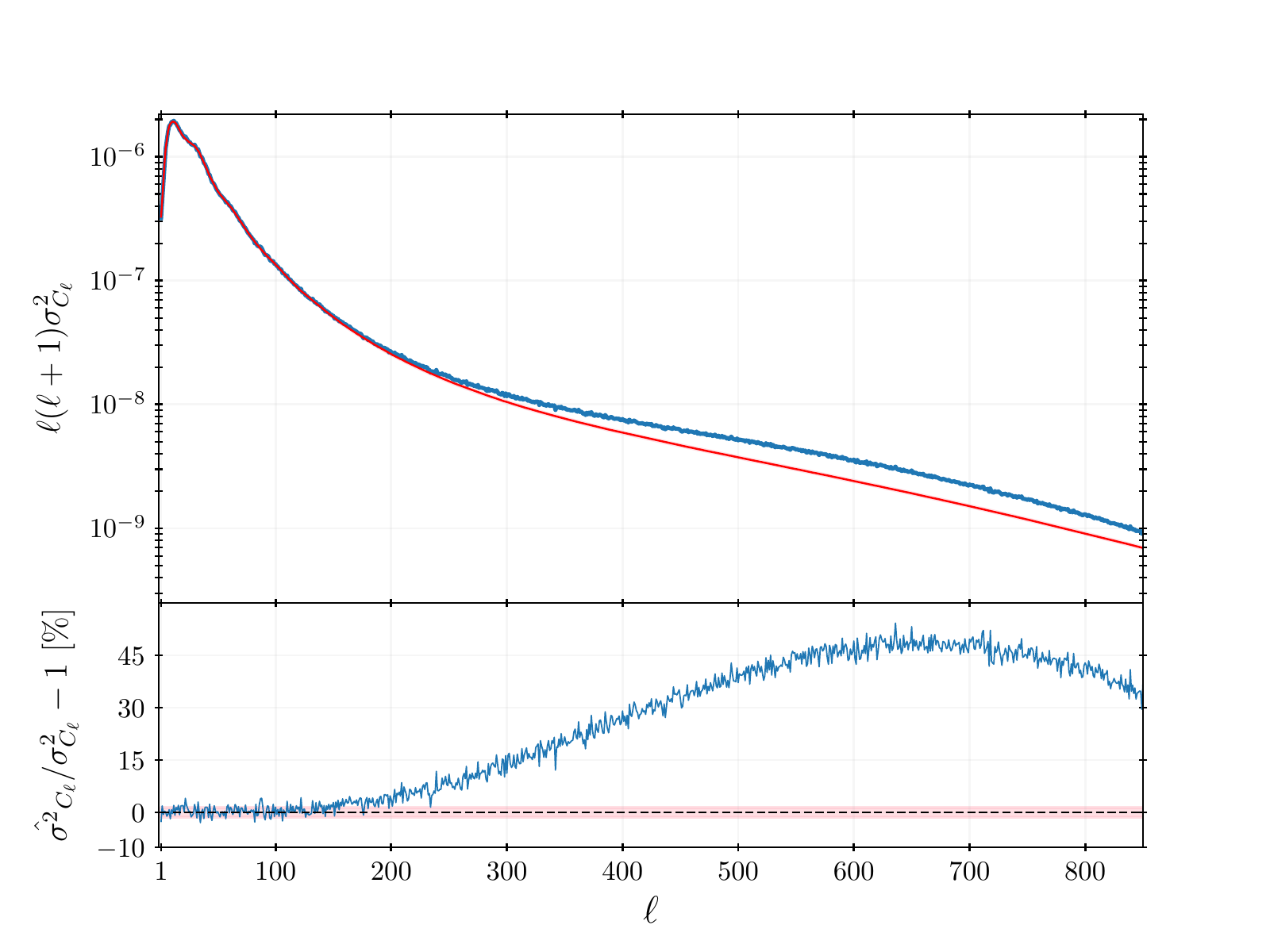}
\caption{{\it Top :} Measured diagonal of the covariance matrix (blue curve) over $N=10000$ realisations of different light cones. The red curve represent the associated prediction in the case of a gaussian field with errors computed using equation \ref{varc}. Here we keep the SN effect in the measures and include it in the prediction. The spherical Nyquist mode is situated around $\ell_N\sim 650$. {\it Bottom :} Relative difference in percent following the same color code.}
\label{diagcovcls}
\end{figure}

We generate $N=10000$ realisations and measure the angular power spectrum in each of them, in order to finally estimate the covariance matrix in the same way as described in section \ref{sec:sampling}. In Figure \ref{diagcovcls} we represent the diagonal of the covariance matrix, the errors on the covariance matrix elements are computed with equation \ref{varc}. Since in the Gaussian case the relative error expected on the diagonal of the covariance matrix elements is given by $\sqrt{2/(N-1)}$, the interest of using such a large number of realisation is that we expect a $1.4$\% precision on the estimation of the diagonal of the covariance matrix and an absolute precision on the correlation coefficients $r_{ij} = C_{ij}/\sqrt{C_{ii}C_{jj}}$ of roughly $0.02$. In the bottom panel of figure \ref{diagcovcls}  we show the relative deviation between the Gaussian prediction and the measured variance of the angular power spectrum, we see that the maximum of deviation is about $15$\% at $\ell \sim 600$. It appears that deviations from Gaussianity remain small compared to the deviation obtained for the power spectrum covariance matrix (see section \ref{sec:sampling}).

%

\begin{figure}[htbp]
\centering
\includegraphics[width=0.5\textwidth]{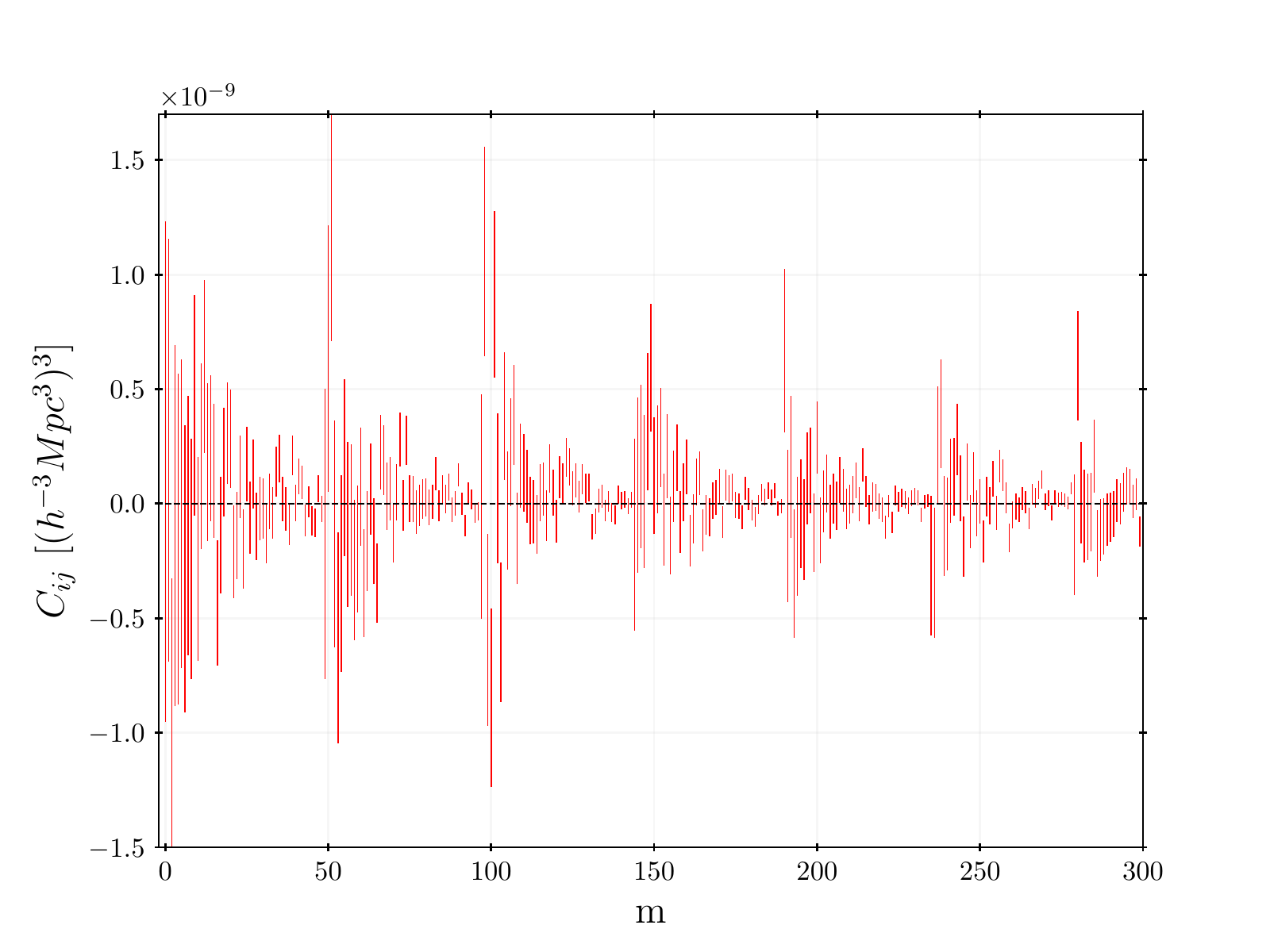}
\caption{The 300 first elements measured of the off-diagonal part of the
   covariance matrix over $n=10000$ realisations of light cone with gaussian errors computed using equation \ref{varc}. The elements
  are labeled by the index m and are ordered column by column of the
  lower half of the matrix without passing by the diagonal, ie. $C_{ij,\ i>j}$. }
\label{offdiagclcovmatrix}
\end{figure}

\begin{figure}[htbp]
\centering
\includegraphics[width=0.55\textwidth]{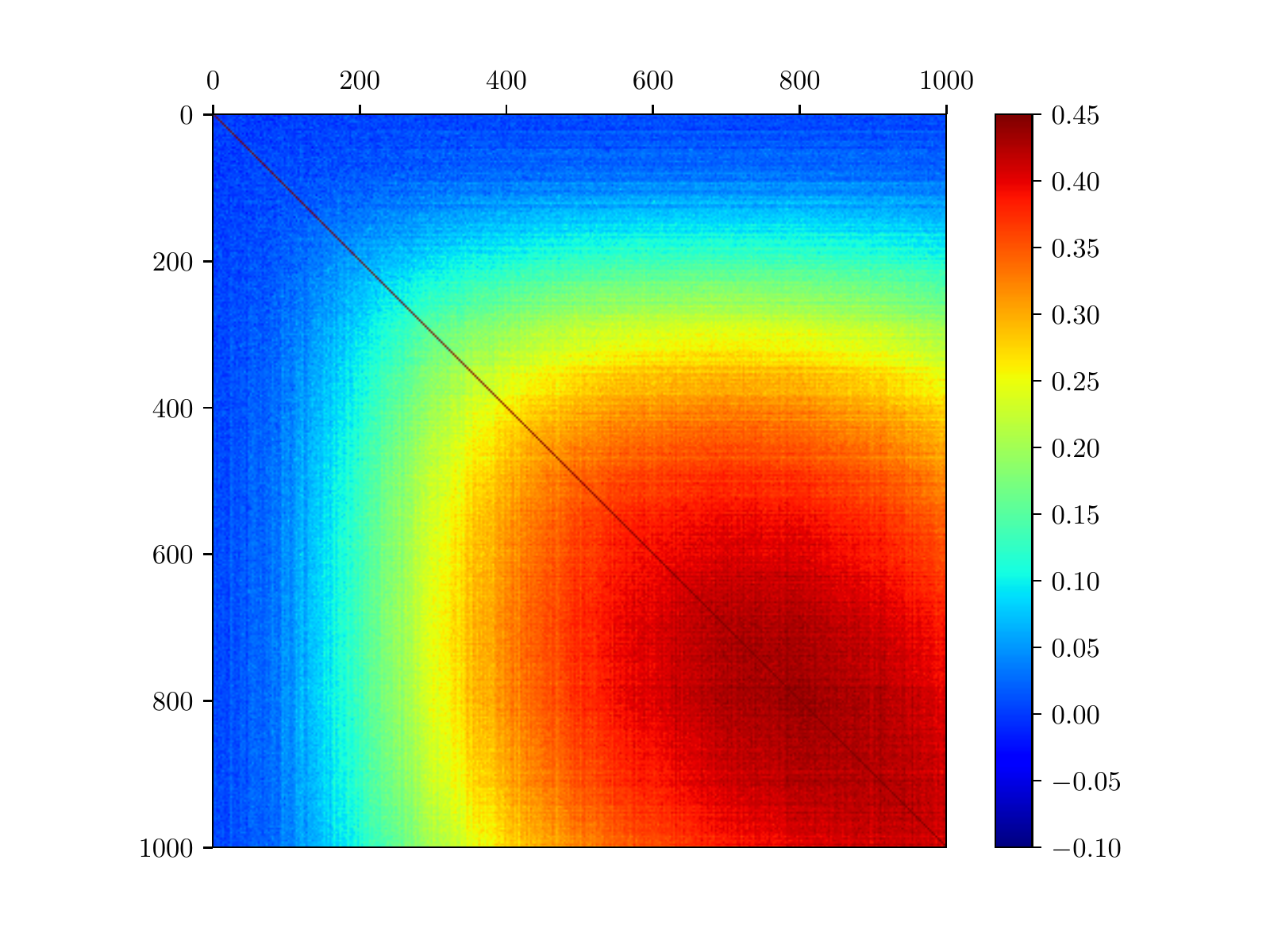}
\caption{Correlation matrix for 10000 realisations of $C_\ell$'s in a simulated universe between redshifts 0.2 and 0.3 and a sampling $N_s=512$. The $(\ell\times \ell')=(1000\times 1000)$ of the matrix are represented here.}
\label{covmatrixcolor}
\end{figure}

In addition, in Figure \ref{offdiagclcovmatrix} we display some off-diagonal covariance elements with their error bars. Despite some fluctuations it is consistant with zero indicating that the covariance matrix is close to be diagonal as expected in the Gaussian case (at least for the 300 firsts elements of the matrix by counting them following the description in caption). In order to make sure that this is indeed the case, in Figure \ref{covmatrixcolor} we show the correlation coefficients $r_{ij}=C_{ij}/\sqrt{C_{ii}C_{jj}}$, we see that the matrix is close to be diagonal only considering $\ell < 200$. It therefore confirms that projecting a thick redshift shell onto the sky tends to turn more Gaussian the density field. This is coherent with what one would naively expect from the central limit theorem, since the projection is made by summing over many values of a non-Gaussian field with some weights its appears that the resulting distribution should tend to a Gaussian as the volume of the projection increases. However, for large $\ell$-values we measure a significant amount of correlation typically of order $10$\% reaching $30$\% at $\ell \sim 600$.


%% file: conclusion.tex
\section*{Conclusion}
\label{conclusion}


In this paper we refined a known process allowing to generate a non-Gaussian density field with a given \pdf and power spectrum. 

We first pointed out the main current mathematical issue arising when one wants to generate a density field with a cut-off scale by filtering its power spectrum. Indeed, we demonstrated that the power spectrum of the Gaussian field that will eventually be transformed into a non-Gaussian is likely to be undefined (i.e. with negative values) on some bandwidth. Despite the fact that, we stressed that there is in principle no way of sorting mathematically this problem we have shown that a simple criterion allows to get around: for a Gaussian filtering, one needs to take a spatial sampling rate $a$ which is larger than twice the cutoff scale $R_f$.



We demonstrated that taking into account aliasing at the stage of generating the density field in Fourier space is of paramount importance in order to maintain the output power spectrum under control. In addition, we have shown that without imposing an explicit cutoff scale, at the stage of producing a catalogue with a local Poisson process, we introduce an effective filtering of the density field which can be predicted with a subpercent level accuracy. Regarding the Poisson sampling, we proposed a natural extension of the usual Top-Hat method consisting in populating the cubical cells uniformly with objects: we can linearly interpolate the density field between nodes and populate the cells with a probability distribution following the interpolated density field. The interest of this extension is that it allows to get closer to the ideal power spectrum by strongly decreasing the amplitude of aliasing.

    
Regarding the density field, we have shown that one can predict in a perturbative way the expected bi-spectrum and tri-spectrum and provided an analytical approximation allowing to predict the variance of the power spectrum estimated for the non-Gaussian density field. This allowed us to check that the statistical behaviour of our method was going in the expected direction.

At the end of section \ref{sec:catalog} we discussed two different methods to build a light cone out of our catalogues. 
The {\it shell method} can be used no matter if the evolution of the power spectrum is linear or not but involves a large number of redshift shell which is time consuming.   The other is much faster,  the {\it cell method} suits particularly well when the power spectrum evolves linearly but does not allow to keep a perfect control on the power spectrum when it presents a non-linear evolution. 

Finally, we presented a possible application of this kind of Monte-Carlo catalogues of objects to tomographic analysis. We have shown that the estimated angular power spectrum is in agreement at the percent level with the expected one. This validated the {\it shell} and {\it cell methods} used to build the light cones. Thanks to the numerical efficiency of this Monte-Carlo we could generate $10000$ realisations allowing to estimate the covariance matrix elements with a percent accuracy. Despite the reduction of non-Gaussianity involved in the projection of the catalogue on the sky, we are still able to detect on small scales clear signature coming from the fact that the catalogues have been generated out of a non-Gaussian density field. 


Such a Monte-Carlo method might be useful in 
investigating the dependance of the covariance matrix on the cosmological parameter. As recently done by \citet{Lippich:2018wrx} and \citet{Blot:2018oxk}, in a future work we plan
to compare the covariance matrix obtained with this 
method to the one estimated from cosmological N-body simulations
and will include the treatment of redshift space distortions.  


%% file: appendix_LN.tex
\section{Some properties of the LN field}\label{appendix_LN}

Let X follow a Gaussian distribution  $X \sim {\cal N}(\mu,\sigma^2)$
then $Y=e^X$ follows a LN distribution.
For simplicity we consider in the following that the Gaussian has a
null mean $\mu=0$.
Its moments can be immediately computed
\begin{align}
  \E{y^k}=& \int_0^\infty y^k f_Y(y) dy=\int_0^\infty e^{k \ln y}
  f_Y(y) dy \nonumber \\
  =& \int_{-\infty}^{+\infty} e^{k x} {\cal N}(x;0,\sigma^2) dx \nonumber \\
= & e^{k^2 \sigma^2/2}.
\end{align}
In particular 
\begin{align}
\label{eq:mom2}
  \E{y}&=e^{\sigma^2/2} \\
  \Var{y}&=e^{\sigma^2}(e^{\sigma^2}-1)
\end{align}

The idea for cosmology is to ensure a positive \textit{energy density}
(noted $\rho$ in the following)
by transforming a Gaussian \textit{density contrast} ($\delta^g$) into
\begin{align}
\label{eq:deltarho}
  \rho_{LN}(x)=e^{\delta^g(x)}
\end{align}

One recovers the LN contrast using Eq.\ref{eq:mom2}
\begin{align}
  \delta_{LN}&=\dfrac{\rho_{LN}}{\E{\rho_{LN}}}-1 = e^{\delta^g-\tfrac{\sigma^2}{2}} -1.
\end{align}
This is a linear transformation of the pure LN distribution $e^{\delta^g}$ so
we can compute immediately its first 2 moments:
\begin{align}
\E{\delta_{LN}}&=0\\
\label{eq:varlog}
\Var{\delta_{LN}}& =[e^{-\sigma^2/2}]^2 \Var{\rho_{LN}}=e^{\sigma^2}-1
\end{align}

The random field is created by considering it a function of
spatial coordinates $x_i$ and from now on we will use the shorthand
$\delta_i=\delta(x_i)$ or $\rho_i=\rho(x_i)$, dropping the "LN" subscript.
Its autocorrelation function, assuming isotropy reads
\begin{align}
  \xi(r)=\E{\delta_1 \delta_2}=\dfrac{\E{(\rho_1-\bar \rho)(\rho_2-\bar
    \rho)}}{\E{\rho}^2}=\dfrac{\E{\rho_1\rho_2}}{\E{\rho}^2}-1
\end{align}
Calling $f_2(\rho_1,\rho_2)$ the 2D density distribution of the LN
\textit{energy density} random field, probability conservation yields
\begin{align}
  f_2(\rho_1,\rho_2) d\rho_1 d\rho_2={\mathcal
  N}(\delta^g_1,\delta^g_2; \mat C) d \delta^g_1 d\delta^g_2
\end{align}

The covariance matrix of the Gaussian field beeing
\begin{align}
  \mat{C}=\E{\delta^g_1 \delta^g_2}=
  \begin{pmatrix}
    \sigma^2 & \xi^g \\
    \xi^g & \sigma^2
  \end{pmatrix}
\end{align}
One can then compute its 2-point function in a way similar to moments
\begin{align}
  \E{\rho_1 \rho_2}&=\iint e^{\ln \rho_1} e^{\ln \rho_2} f_2(\rho_1,\rho_2) d\rho_1 d\rho_2 \\
&=\iint e^{\delta^g_1} e^{\delta^g_2} {\mathcal N}(\delta^g_1,\delta^g_2;\mat C) d\delta^g_1 d\delta^g_2\\
&=e^{\sigma^2+\xi}.
\end{align}
The last line can be obtained from a direct computation or
recalling that the generative functional of a multi-dimensional
Gaussian is
\begin{align}
  \E{e^{xt}}=e^{\tfrac{1}{2}t^T\mat{C}t},
\end{align}
$x, t$ representing vectors.

Finally, using Eqs.\ref{eq:deltarho} and \ref{eq:mom2}, one obtains
for the \textit{contrast density} of the LN field the beautiful result that
\begin{align}
  \xi_{LN}(r)=e^{\xi^g(r)}-1
\end{align}

One may check that the variance ($\xi(r=0)$) indeed follows Eq.\ref{eq:varlog}.

%% file: appendix_Mehler.tex
\section{The Mehler formalism}\label{app:mehler}
The Mehler transform for bivariate distributions is not a well
known tool, while it is particularly convenient to ease computations
of 2D integrals involving Gaussian distributions
as was demonstrated in \citet{Simpson13} or more recently in \citet{Bel:2016}.

Let $(X_1,X_2)$ follow a central bivariate  distribution
\begin{align}
  (X_1,X_2)\sim\Norm_2(0,\mat{\Sigma}),
\end{align}
with a covariance matrix
\begin{align}
  \mat{\Sigma}=
  \begin{pmatrix}
    1 & \xi_X\\
    \xi_X & 1
  \end{pmatrix}.
\end{align}
For convenience we restrict the variance term to 1 , so that the
covariance term  $\xi_\nu$ is the correlation coefficient, and we will show at
the end of this appendix how to treat the general case.
By denoting, in loose notation, $\Norm(x)$ as the 1D normal
distribution, the transform reads
 
\begin{align}
  \Norm_2(x_1,x_2)=\Norm(x_1)\Norm(x_2) \sum_{n=0}^{\infty}
  H_n(x_1) H_n(x_2) \dfrac{\xi_X^n}{n!},
\end{align}

where $H_n$ are the
(probabilistic) Hermite polynomials which are orthogonal wrt to the Gaussian measure
\begin{align}
  \int_{-\infty}^{+\infty} H_n(x) H_m(x) \Norm(x) dx=n! \delta_{n m}.
\end{align}

The interest here is that when applying some local transform to a
Gaussian field $Y=\lcal(X)$ the covariance of the transformed field
becomes
\begin{align}
\label{eq:xitrans}
  \xi_Y=\E{y_1y_2}&=\int_{-\infty}^\infty\int_{-\infty}^\infty y_1 y_2 f_Y(y_1,y_2) dy_1 dy_2 \nn \\
& =\int_{-\infty}^\infty\int_{-\infty}^\infty \lcal(x_1)\lcal(x_2) \Norm_2(x_1,x_2) dx_1 dx_2.
\end{align}

Then if we decompose the local field onto the Hermite polynomials
\begin{align}
  \lcal(x)=\sum_{n=0}^\infty c_n H_n(x),
\end{align}
and use the orthogonality properties, one obtains the simple expansion
\begin{align}
\label{eq:series}
  \xi_Y&=\sum_{n=0}^\infty n! c_n^2 \xi_G^n,
\end{align}
where 
\begin{align}
\label{eq:cn}
  c_n=\dfrac{1}{n!}\int_{-\infty}^{+\infty} \lcal(x) H_n(x) \Norm(x) dx.
\end{align}

An important point to notice is that \textit{all the coefficients in
  the expansion are  positive}. Compare this to the series expansion of the inverse
log-transform (Eq. \ref{eq:lnxi}).
One sees immediately that a field with a $\ln(1+\xi_X$) covariance
cannot be obtained from a Gaussian one.

Let us now reconsider the classical log-normal field (but with
$\sigma=1$). The local transform reads
\begin{align}
  \lcal(x)&=e^{x-1/2} -1.
\end{align}
Then
\begin{align}
  c_n&=\dfrac{1}{n!}\int_{-\infty}^{+\infty} (e^{x-1/2}-1)
  \Norm(x) H_n(x) dx \nn \\
  & =\dfrac{1}{n!} \left[ \int_{-\infty}^{+\infty} \Norm(x-1) H_n(x) dx -
  \int_{-\infty}^{+\infty} \Norm(x) H_n(x) dx \right] \nn \\
&=\dfrac{1}{n!}[1-\delta_{n0}].
\end{align}
where we used the $\Norm(x)$ expression, $H_0(x)=1$ and $H_n(x+1)=\displaystyle{\sum_{k=0}^n}\binom{n}{k}H_{n-k}(x)$.

From Eq.\ref{eq:series} the autocorrelation of the LN field is
\begin{align}
\label{eq:tropfort}
 \xi_Y=\sum_{n=1}^\infty \dfrac{\xi_G}{n!} = e^{\xi_G}-1,
\end{align}
in agreement with the more classical way to derive it shown in 
Appendix \ref{appendix_LN}.

While unnecessary in the LN case, such an approach is very powerful
in computing numerically the auto-correlation of any transformed
Gaussian field.

When the Gaussian field does not have a unit variance
$\sigma^2=\xi_X(0)\ne 1$ which is generally the case,
one works with rescaled variables leading
to

\begin{align}
  \xi_Y&=\sum_{n=0}^\infty n! c_n^2 \left(
         \dfrac{\xi_\nu}{\sigma^2}\right)^n, \\
  c_n&=\dfrac{1}{n!}\int_{-\infty}^{+\infty} \lcal(\sigma x) H_n(x) \Norm(x) dx.
\end{align}
Then using the more general local transform discussed in Appendix
\ref{appendix_LN}, 
 \begin{align}
  \lcal(x)&=e^{x-\sigma^2/2} -1,
\end{align}
one recovers Eq.\ref{eq:tropfort}.

%% file: appendix_trispectre.tex
\onecolumn
\section{Higher order correlation functions}
\label{appendix_trispectre}

In this appendix we show how to predict in a perturbative way the bi-spectrum and tri-spectrum a non-gaussian density field generated from the local transformation of a Gaussian field. 

Let us consider a density field $\epsilon(\vec x)$ in configuration space. We can therefore define its Fourier transform as

\begin{equation}
\epsilon_{\vec k}  = \fcal\left [ \epsilon(\vec x)\right ] \equiv \frac{1}{(2\pi)^3}\int \epsilon(\vec x) e^{-\vec k \cdot \vec x}\dif^3\vec x.
\label{fourier}
\end{equation}
As explained is section \ref{sec:pipeline} we generate a Gaussian random field in Fourier space (assuming a power spectrum), we inverse Fourier transform it to get its analog in configuration space. We further apply a local transform $\lcal$ to map the Gaussian field into a stochastic field that is characterised by a target \pdf. Thus, the $N$-point moments can be in principle predicted as soon as the local transform and the target power spectrum have been specified. 

Be $\nu$ a stochastic field following a centered ($\langle\nu\rangle=0$) reduced ($\sigma_\nu^2\equiv\langle\nu^2\rangle_c=1$) Gaussian distribution. From a realisation of this field, one can generate a non-Gaussian density field $\delta$ by applying a local mapping $\lcal$ between the two, hence

\begin{equation}
\delta = \lcal(\nu).
\label{local}
\end{equation}
Without lake of generality,  one can express the $N$-point moments of the transformed density field with respect to the $2$-point correlation of the Gaussian field as 

\begin{equation}
\langle\delta_1...\delta_N\rangle =\int \lcal(\nu_1)...\lcal(\nu_N) \mathcal{B}^{(N)}(\vec \nu , C_\nu)\dif \nu_1...\dif \nu_N,
\label{npointmom}
\end{equation}
where $\mathcal{B}^{(N)}$ is a $N$-variate Gaussian distribution with a $N\times N$ covariance matrix $C_\nu$ and sub-indexes are referring to positions $\delta_1\equiv \delta(\vec x_1)$. In practice the computation of equation \ref{npointmom} can be numerically expensive, however as shown in appendix \ref{app:mehler} it can be efficiently computed thanks to the Mehler expansion, at least in the case of the $2$-point moment (see equation \ref{eq:series}). Assuming that in the local transform $\lcal$ the amplitude of the coefficients of its Hermite transform (see equation \ref{eq:cn}) is decreasing with the order $n$, it offers the possibility of ordering the various contributions  to the total moment.  

In order to evaluate equation \ref{npointmom} one can use extensions of the Mehler formula \citep{Carlitz:1970}, for example the third order leads to

\begin{equation}
{\mathcal B}^{(3)}(\vec \nu, C_\nu) = \sum_{m,n,p}^{\infty} \frac{H_{n+p}(\nu_1)}{m!}\frac{H_{p+m}(\nu_2)}{n!}\frac{H_{n+m}(\nu_3)}{p!}\xi_{23}^m\xi_{13}^n\xi_{12}^p G^{(3)}(\vec \nu) 
\label{b3}
\end{equation}
where the correlation functions $\xi_{12}$, $\xi_{13}$ and $\xi_{23}$ are the three off-diagonal elements of the covariance matrix $C_\nu$ and the function $G^{N}$ is defined as an $N$-variate Gaussian distribution with a diagonal covariance matrix which values are all set to unity

\begin{equation}
G^{(N)}(\vec \nu) \equiv \frac{1}{(2\pi)^{N/2}}e^{-\frac{1}{2}(\nu_1^2+...+\nu_N^2)}.
\label{gaussdiag}
\end{equation}
At fourth order ($N=4$), the $4$-variate Gaussian can be expressed as 

\begin{equation}
{\mathcal B}^{(4)}(\vec \nu, C_\nu) =\sum_{l,m,n,o,p,q}^{\infty} \frac{H_{l+m+n}(\nu_1) H_{l+o+p}(\nu_2) H_{m+o+q}(\nu_3) H_{n+p+q}(\nu_4)}{l!m!n!o!p!q!}\xi_{12}^l\xi_{13}^m \xi_{14}^n \xi_{23}^o \xi_{24}^p \xi_{34}^q G^{(4)}(\vec \nu)
\label{b4}
\end{equation}
where again $\xi_{12}$, $\xi_{13}$,  $\xi_{14}$, $\xi_{23}$, $\xi_{24}$ and $\xi_{34}$ are the $6$ off-diagonal elements of the covariance matrix $C_\nu$. By replacing  equations \ref{b3} and \ref{b4} in expression \ref{npointmom} one can integrate over the $N$ variables $\nu_1$ to $\nu_N$ and express the $3$- and $4$-points moments as a sum over product of the two point correlation function of the Gaussian field

\begin{equation}
\langle\delta_1\delta_2\delta_3\rangle = \sum_{m,n,p}^{\infty} \frac{(n+p)!(p+m)!(n+m)!}{m!n!p!}c_{n+p}c_{p+m}c_{n+m}\xi_{23}^m\xi_{13}^n\xi_{12}^p,
\label{mom3}
\end{equation}
and

%

\begin{equation}
\langle\delta_1\delta_2\delta_3\delta_4\rangle = \sum_{l,m,n,o,p,q}^{\infty} \textstyle \frac{(l+m+n)!(l+o+p)!(m+o+q)!(n+p+q)!}{l!m!n!o!p!q!}c_{l+m+n}c_{l+o+p}c_{m+o+q}c_{n+p+q}\xi_{12}^l\xi_{13}^m \xi_{14}^n \xi_{23}^o \xi_{24}^p \xi_{34}^q,
\label{mom4}
\end{equation}
the coefficients $c_i$ are still the coefficients of the Hermite expansion defined by equation \ref{eq:cn}. Equations \ref{mom3} and \ref{mom4} are particularly useful when one wants to evaluate the $3$- and $4$-point correlation functions of the density field $\delta$ or their Fourier counterparts the bi-spectrum and tri-spectrum. 

Let us express, first, the power spectrum of the density field $\delta$ with respect to the power spectrum of the Gaussian field $\nu$. By Fourier transforming equation \ref{eq:series} one can obtain 

\begin{equation}
P_\delta(k) = c_1^2 P(k) + \sum_{n=2}^\infty n!c_n^2P^{(n)}(k),
\label{ploops}
\end{equation}
where the $P^{(n)}(k)$ represent what we will call loop corrections of order $n-1$ and are defined as $P^{(n)}(k) \equiv \fcal [\xi_{\nu}^n]$. The leading order or tree-level contribution is given by $c_1^2 P(k)$ which is just a change in amplitude of the power spectrum of the Gaussian field. It represents the change of the power spectrum one would expect if the local transformation $\lcal$ was linear. 

We now express the $3$-point correlation function $\zeta_{\delta,123}\equiv \langle \delta_1\delta_2\delta_3\rangle_c$, dropping terms higher then  $1$-loop corrections one would obtain 

%

\begin{equation}
\zeta_{\delta, 123} \simeq  2c_2c_1^2\left [ \xi_{12}\xi_{13} + \xi_{12}\xi_{23} + \xi_{13}\xi_{23} \right ] + 6c_3c_1c_2 \left [  \xi_{12}\xi_{13}^2 + \xi_{12}\xi_{23}^2 + \xi_{13}\xi_{23}^2 + \xi_{12}^2\xi_{13} + \xi_{12}^2\xi_{23} + \xi_{13}^2\xi_{23}    \right ] + 8c_2^3\xi_{12}\xi_{13}\xi_{23}. 
\label{3pcf}
\end{equation}
Taking the Fourier transform of the above equation \ref{3pcf} one can obtain the expression of the bi-spectrum of the density field as 

\begin{eqnarray}
B_\delta(k_1, k_2) & \simeq & 2c_2c_1^2\left [ P(k_1)P(k_2) + P(k_1)P(k_{12}) + P(k_2)P(k_{12}) \right ] + \nonumber \\
                          &             & 6c_3c_1c_2 \left [ P(k_1)P^{(2)}(k_2) + P(k_1)P^{(2)}(k_{12}) + P(k_2)P^{(2)}(k_{12})  + \right . \nonumber  \\
                          &             & \left .      P^{(2)}(k_1)P(k_2) + P^{(2)}(k_1)P(k_{12}) + P^{(2)}(k_2)P(k_{12})   \right ] + 8c_2^3B^{(3)}(k_1, k_2), 
\label{bispec}
\end{eqnarray}
where we use the short-hand notations $k_i = \vec k_i $, $k_{ij} = |\vec k_i + \vec k_j |$  and

\begin{equation}
B^{(3)}(k_1, k_2) \equiv \frac{1}{(2\pi)^6} \int \xi (r) \xi (s) \xi (| \vec s - \vec r |) e^{-i\vec k_1\cdot\vec r - i \vec k_2\cdot\vec s}\dif^3\vec r \dif^3\vec s,
\label{bi3}
\end{equation}
which can also be expressed in terms of a triple product of the power spectrum at different wave modes

\begin{equation}
B^{(3)}(k_1, k_2) = \int P(q) P(|\vec q+\vec k_1|) P(|\vec q-\vec k_2|) \dif^3\vec q . 
\label{bi3pk}
\end{equation}
In the very same way one can also express the $1$-loop prediction of the tri-spectrum, we need to start from the four-point correlation function $\eta_{\delta,1234}\equiv \langle\delta_1\delta_2\delta_3\delta_4\rangle_c = \langle\delta_1\delta_2\delta_3\delta_4\rangle - \xi_{\delta, 12}\xi_{\delta, 34} - \xi_{\delta, 13}\xi_{\delta, 24} - \xi_{\delta, 14}\xi_{\delta, 23}$ \citep{Fry84b}, where we need to express the products of $2$-point correlation functions at fourth order, it follows 

%

\begin{equation}
\xi_{\delta, 12}\xi_{\delta, 34} \simeq c_1^4 \xi_{12}\xi_{34} + 2c_2^2c_1^2\left [ \xi_{12}^2\xi_{34} + \xi_{12}\xi_{34}^2\right ] + 4 c_2^4\xi_{12}^2\xi_{34}^2 + 6c_3^2 c_1^2 \left [ \xi_{12}^3\xi_{34} + \xi_{12}\xi_{34}^3 \right ]. 
\label{xiprod}
\end{equation}
Keeping terms of order lower or equal to four in terms of $\xi$ in equation \ref{mom4} and subtracting permutations of equation \ref{xiprod}  one can obtain the $1$-loop expression of the four-point correlation function 

\begin{eqnarray}
\eta_{\delta, 1234} & \simeq & 6c_3c_1^3 \left [ \xi_{12}\xi_{13}\xi_{14}+ 3\; {\rm perm.} \right ] + 4c_2^2c_1^2 \left [ \xi_{12}\xi_{23}\xi_{34}+ 11\; {\rm perm.} \right ] + \nonumber \\
                          &             & 18c_1^2c_3^2 \left [ \xi_{12}\xi_{13}^2\xi_{34} + 11\; {\rm perm.}  \right ]+ \nonumber \\
                          &             & 12c_3c_2^2c_1 \left [ \xi_{12}\xi_{13}\xi_{34} ( \xi_{12} + \xi_{34} ) + 11\; {\rm perm.}  \right ] + \nonumber \\ 
                          &             & 24 c_4c_2c_1^2 \left [ \xi_{12}\xi_{13}\xi_{14} ( \xi_{12} + \xi_{13} + \xi_{14} ) + 3\; {\rm perm.}  \right ] + \nonumber \\ 
                          &             & 24c_1c_2^2c_3\left [ \xi_{12}\xi_{34} ( \xi_{13}\xi_{14} + \xi_{23}\xi_{24} +\xi_{13}\xi_{23}+\xi_{14}\xi_{24} ) + 2\; {\rm perm.}  \right ]+ \nonumber \\
                          &             & 16 c_2^4\left [ \xi_{12}\xi_{14}\xi_{23}\xi_{34} + 2\; {\rm perm.}  \right ]. 
\end{eqnarray}
In order to recover the correct permutations, one has to notice that $\xi_{12}$ and $\xi_{34}$, $\xi_{13}$ and $\xi_{24}$, $\xi_{14}$ and $\xi_{23}$ can be interchanged without modification of the coefficients in front, thus in the second line we have four permutations involving the product $\xi_{12}\xi_{34}$ and we can iterate three times by taking the mentioned specific pairs ($\xi_{12}\xi_{34}$,  $\xi_{13}\xi_{24}$ and $\xi_{14}\xi_{23}$). The above expression can be transformed into the tri-spectrum by just taking its Fourier transform, it reads

\begin{eqnarray}
T_\delta(k_1, k_2, k_3) & \simeq & 4c_2^2c_1^2 \left \{ P(k_1)P(k_2)\left [ P(k_{13}) + P(k_{14})\right ] + 5\; {\rm perm.}  \right \} + \nonumber \\ 
                                 &             & 6c_3c_1^3 \left \{ P(k_1)P(k_2)P(k_3) + 3\; {\rm perm.}  \right \} + \nonumber \\ 
                                 &             & 18c_1^2c_3^2 \left \{ P^{(2)}(k_{12})P(k_1)P(k_3) + 11\; {\rm perm.}  \right \} + \nonumber \\ 
                                 &             & 12c_3c_2^2c_1 \left \{ P(k_{12})\left [ P^{(2)}(k_1)P(k_3) +P(k_1)P^{(2)}(k_3) \right ]+ 11\; {\rm perm.}  \right \} + \nonumber \\ 
                                 &             & 24c_4c_2c_1^2 \left \{ P(k_1)P(k_2)P^{(2)}(k_3)+ 11\; {\rm perm.}  \right \}  + \nonumber \\
                                 &             & 24c_3c_2^2c_1 \left \{ P(k_1)B^{(3)}(k_2, k_{3}) + 11\; {\rm perm.}  \right \}  + \nonumber \\
                                 &             & 16c_2^4 \left \{ T^{(4)}(k_1, k_{2}, k_{14})+T^{(4)}(k_1, k_{3}, k_{12}) +  T^{(4)}(k_1, k_{4}, k_{13})  \right \} ,  \label{trid}
\end{eqnarray}
where we define the fourth order tri-spectrum as

\begin{equation}
T^{(4)}(k_1, k_2, k_3) \equiv \int P(q) P(|\vec q+\vec k_1|) P(|\vec q-\vec k_2|) P(|\vec q+\vec k_3|) \dif^3\vec q . 
\label{tri4}
\end{equation}
Of course, the practical evaluation of all the terms in equation \ref{trid} is not easy to get, however in order to predict the covariance matrix we only need specific configuration of the tri-spectrum  the problem can simplified when trying to predict only the diagonal contribution ($k_i = k_j$). This has been already considered in literature \citep{SZ&H99} at the tree-level and we obtain the same expression which is

\begin{equation}
\bar T(k_i, k_i) \sim  8 c_1^2 \left \{ 4c_2^2 + 3c_3c_1 \right \} P^3(k_i), 
\label{tridbar}
\end{equation}
where as in \citet{SZ&H99} we approximate the angular averages the power spectrum of the sum of two wave modes as being equal to the power spectrum evaluated at half the modulus of the two wave modes. A simple extension of the previous result can be obtained by neglecting the $B^{(3)}$ and $T^{(4)}$ terms in equation \ref{tri4}, one would  get 

\begin{equation}
\bar T(k_i, k_i) \sim  8 c_1^2 \left \{ 4c_2^2 + 3c_3c_1 \right \} P^3(k_i)  + 24 \left \{ 3c_1^2c_3^2 + 4c_1c_2^2c_3+ 12c_1^2c_2c_4 \right \} P^2(k_i)P^{(2)}(k_i) + 144 c_1^2c_3^2P^{(2)}(0)P^2(k_i), 
\end{equation}
which can be used to control the statistical behaviour of our Monte-Carlo density fields. 


\twocolumn

%% file: Article1.bbl
\begin{thebibliography}{40}
\expandafter\ifx\csname natexlab\endcsname\relax\def\natexlab#1{#1}\fi

\bibitem[{{Adler}(1981)}]{Adler:1981}
{Adler}, R.~J. 1981, The Geometry of Random Fields (Wiley)

\bibitem[{{Agrawal} {et~al.}(2017){Agrawal}, {Makiya}, {Chiang}, {Jeong},
  {Saito}, \& {Komatsu}}]{Agrawal:2017}
{Agrawal}, A., {Makiya}, R., {Chiang}, C.-T., {et~al.} 2017, Journal of
  Cosmology and Astro-Particle Physics, 2017, 003

\bibitem[{{Anderson}(1984)}]{Anderson}
{Anderson}, T.~W. 1984, {An introduction to multivariate statistical analysis,
  second edition} (Wiley series in probability and mathematical statistics)

\bibitem[{{Asorey} {et~al.}(2012){Asorey}, {Crocce}, {Gazta{\~n}aga}, \&
  {Lewis}}]{Asorey12}
{Asorey}, J., {Crocce}, M., {Gazta{\~n}aga}, E., \& {Lewis}, A. 2012, \mnras,
  427, 1891

\bibitem[{{Bel} {et~al.}(2016){Bel}, {Branchini}, {Di Porto}, {Cucciati},
  {Granett}, {Iovino}, {de la Torre}, {Marinoni}, {Guzzo}, {Moscardini},
  {Cappi}, {Abbas}, {Adami}, {Arnouts}, {Bolzonella}, {Bottini}, {Coupon},
  {Davidzon}, {De Lucia}, {Fritz}, {Franzetti}, {Fumana}, {Garilli}, {Ilbert},
  {Krywult}, {Le Brun}, {Le F{\`e}vre}, {Maccagni}, {Ma{\l}ek}, {Marulli},
  {McCracken}, {Paioro}, {Polletta}, {Pollo}, {Schlagenhaufer}, {Scodeggio},
  {Tasca}, {Tojeiro}, {Vergani}, {Zanichelli}, {Burden}, {Marchetti},
  {Mellier}, {Nichol}, {Peacock}, {Percival}, {Phleps}, \& {Wolk}}]{Bel:2016}
{Bel}, J., {Branchini}, E., {Di Porto}, C., {et~al.} 2016, \aap, 588, A51

\bibitem[{Blas {et~al.}(2011)Blas, Lesgourgues, \& Tram}]{Blas:2011rf}
Blas, D., Lesgourgues, J., \& Tram, T. 2011, JCAP, 1107, 034

\bibitem[{Blot {et~al.}(2019)}]{Blot:2018oxk}
Blot, L. {et~al.} 2019, Mon. Not. Roy. Astron. Soc., 485, 2806

\bibitem[{{Bonvin} \& {Durrer}(2011)}]{B&D11}
{Bonvin}, C. \& {Durrer}, R. 2011, \prd, 84, 063505

\bibitem[{{Cai} \& {Bernstein}(2012)}]{C&B12}
{Cai}, Y.-C. \& {Bernstein}, G. 2012, \mnras, 422, 1045

\bibitem[{{Campagne} {et~al.}(2018){Campagne}, {Neveu}, \&
  {Plaszczynski}}]{Angpow:2018}
{Campagne}, J.-E., {Neveu}, \& {Plaszczynski}, S. 2018, {AngPow: Fast
  computation of accurate tomographic power spectra}, Astrophysics Source Code
  Library

\bibitem[{{Carlitz}(1970)}]{Carlitz:1970}
{Carlitz}, L. 1970, Collect. Math., 21, 117

\bibitem[{{Chiang} {et~al.}(2013){Chiang}, {Wullstein}, {Jeong}, {Komatsu},
  {Blanc}, {Ciardullo}, {Drory}, {Fabricius}, {Finkelstein}, {Gebhardt},
  {Gronwall}, {Hagen}, {Hill}, {Jee}, {Jogee}, {Land riau}, {Mentuch Cooper},
  {Schneider}, \& {Tuttle}}]{Chiang:2013}
{Chiang}, C.-T., {Wullstein}, P., {Jeong}, D., {et~al.} 2013, Journal of
  Cosmology and Astro-Particle Physics, 2013, 030

\bibitem[{{Clerkin} {et~al.}(2017){Clerkin}, {Kirk}, {Manera}, {Lahav},
  {Abdalla}, {Amara}, {Bacon}, {Chang}, {Gazta{\~n}aga}, {Hawken}, {Jain},
  {Joachimi}, {Vikram}, {Abbott}, {Allam}, {Armstrong}, {Benoit-L{\'e}vy},
  {Bernstein}, {Bernstein}, {Bertin}, {Brooks}, {Burke}, {Rosell}, {Carrasco
  Kind}, {Crocce}, {Cunha}, {D'Andrea}, {da Costa}, {Desai}, {Diehl},
  {Dietrich}, {Eifler}, {Evrard}, {Flaugher}, {Fosalba}, {Frieman}, {Gerdes},
  {Gruen}, {Gruendl}, {Gutierrez}, {Honscheid}, {James}, {Kent}, {Kuehn},
  {Kuropatkin}, {Lima}, {Melchior}, {Miquel}, {Nord}, {Plazas}, {Romer},
  {Roodman}, {Sanchez}, {Schubnell}, {Sevilla-Noarbe}, {Smith},
  {Soares-Santos}, {Sobreira}, {Suchyta}, {Swanson}, {Tarle}, \&
  {Walker}}]{Clerkin:2017}
{Clerkin}, L., {Kirk}, D., {Manera}, M., {et~al.} 2017, \mnras, 466, 1444

\bibitem[{{Codis} {et~al.}(2016){Codis}, {Pichon}, {Bernardeau}, {Uhlemann}, \&
  {Prunet}}]{Codis:2016}
{Codis}, S., {Pichon}, C., {Bernardeau}, F., {Uhlemann}, C., \& {Prunet}, S.
  2016, \mnras, 460, 1549

\bibitem[{{Coles} \& {Barrow}(1987)}]{Coles:1987}
{Coles}, P. \& {Barrow}, J.~D. 1987, \mnras, 228, 407

\bibitem[{{Coles} \& {Jones}(1991)}]{Coles:1991}
{Coles}, P. \& {Jones}, B. 1991, \mnras, 248, 1

\bibitem[{Coles \& Lucchin(2003)}]{Coles:2003}
Coles, P. \& Lucchin, F. 2003, Cosmology, the Origin and Evolution of Cosmic
  Structure (Wiley)

\bibitem[{{Colombi}(1994)}]{colombi94}
{Colombi}, S. 1994, \apj, 435, 536

\bibitem[{{Crocce} {et~al.}(2013){Crocce}, {Castander}, {Gaztanaga}, {Fosalba},
  \& {Carretero}}]{mice2}
{Crocce}, M., {Castander}, F.~J., {Gaztanaga}, E., {Fosalba}, P., \&
  {Carretero}, J. 2013, ArXiv e-prints

\bibitem[{{Fosalba} {et~al.}(2013){Fosalba}, {Crocce}, {Gaztanaga}, \&
  {Castander}}]{mice1}
{Fosalba}, P., {Crocce}, M., {Gaztanaga}, E., \& {Castander}, F.~J. 2013, ArXiv
  e-prints

\bibitem[{{Fry}(1984{\natexlab{a}})}]{Fry84a}
{Fry}, J.~N. 1984{\natexlab{a}}, \apjl, 277, L5

\bibitem[{{Fry}(1984{\natexlab{b}})}]{Fry84b}
{Fry}, J.~N. 1984{\natexlab{b}}, \apj, 279, 499

\bibitem[{{Gazta{\~n}aga} {et~al.}(2012){Gazta{\~n}aga}, {Eriksen}, {Crocce},
  {Castander}, {Fosalba}, {Marti}, {Miquel}, \& {Cabr{\'e}}}]{Gaztanaga12}
{Gazta{\~n}aga}, E., {Eriksen}, M., {Crocce}, M., {et~al.} 2012, \mnras, 422,
  2904

\bibitem[{{Gazta{\~n}aga} {et~al.}(2000){Gazta{\~n}aga}, {Fosalba}, \&
  {Elizalde}}]{GF&E2000}
{Gazta{\~n}aga}, E., {Fosalba}, P., \& {Elizalde}, E. 2000, \apj, 539, 522

\bibitem[{{G{\'o}rski} {et~al.}(2005){G{\'o}rski}, {Hivon}, {Banday},
  {Wandelt}, {Hansen}, {Reinecke}, \& {Bartelmann}}]{Healpix:2005}
{G{\'o}rski}, K.~M., {Hivon}, E., {Banday}, A.~J., {et~al.} 2005, \apj, 622,
  759

\bibitem[{{Gradshteyn} \& {Ryzhik}(2007)}]{GradshteynRyzhik:2007}
{Gradshteyn}, I.~S. \& {Ryzhik}, I.~M. 2007, {Table of Integrals, Series, and
  Products} (Academic Press)

\bibitem[{{Greiner} \& {En{\ss}lin}(2015)}]{Greiner:2015}
{Greiner}, M. \& {En{\ss}lin}, T.~A. 2015, \aap, 574, A86

\bibitem[{{Hamilton}(2000)}]{Hamilton:2000}
{Hamilton}, A.~J.~S. 2000, \mnras, 312, 257

\bibitem[{{Hockney} \& {Eastwood}(1988)}]{H&E88}
{Hockney}, R.~W. \& {Eastwood}, J.~W. 1988, {Computer simulation using
  particles}

\bibitem[{{Hubble}(1934)}]{Hubble:1934}
{Hubble}, E. 1934, \apj, 79, 8

\bibitem[{{Klypin} {et~al.}(2018){Klypin}, {Prada}, {Betancort-Rijo}, \&
  {Albareti}}]{Klypin:2018}
{Klypin}, A., {Prada}, F., {Betancort-Rijo}, J., \& {Albareti}, F.~D. 2018,
  \mnras, 481, 4588

\bibitem[{{Layzer}(1956)}]{Layzer:1956}
{Layzer}, D. 1956, \aj, 61, 383

\bibitem[{Lippich {et~al.}(2019)}]{Lippich:2018wrx}
Lippich, M. {et~al.} 2019, Mon. Not. Roy. Astron. Soc., 482, 1786

\bibitem[{{Montanari} \& {Durrer}(2012)}]{M&D12}
{Montanari}, F. \& {Durrer}, R. 2012, \prd, 86, 063503

\bibitem[{{Peebles}(1980)}]{Peebles80}
{Peebles}, P.~J.~E. 1980, {The large-scale structure of the universe}

\bibitem[{{Scoccimarro} {et~al.}(1999){Scoccimarro}, {Zaldarriaga}, \&
  {Hui}}]{SZ&H99}
{Scoccimarro}, R., {Zaldarriaga}, M., \& {Hui}, L. 1999, \apj, 527, 1

\bibitem[{{Sefusatti} {et~al.}(2016){Sefusatti}, {Crocce}, {Scoccimarro}, \&
  {Couchman}}]{SCSC16}
{Sefusatti}, E., {Crocce}, M., {Scoccimarro}, R., \& {Couchman}, H.~M.~P. 2016,
  \mnras, 460, 3624

\bibitem[{{Simpson} {et~al.}(2013){Simpson}, {Heavens}, \&
  {Heymans}}]{Simpson13}
{Simpson}, F., {Heavens}, A.~F., \& {Heymans}, C. 2013, \prd, 88, 083510

\bibitem[{{Uhlemann} {et~al.}(2016){Uhlemann}, {Codis}, {Pichon}, {Bernardeau},
  \& {Reimberg}}]{Uhlemann:2016}
{Uhlemann}, C., {Codis}, S., {Pichon}, C., {Bernardeau}, F., \& {Reimberg}, P.
  2016, \mnras, 460, 1529

\bibitem[{Yoglom(1986)}]{Yoglom:1986}
Yoglom, A. 1986, Correlation theory of stationary and related random functions,
  Volume I: Basic Results (Spinger Series in Statistics)

\end{thebibliography}
